\documentclass[tightenlines,eqsecnum,floats,floatfix,aps,prd,notitlepage,nofootinbib,superscriptaddress]{revtex4-2}
\usepackage[utf8]{inputenc}
\usepackage[table]{xcolor}
\usepackage{multirow}
\usepackage{hyperref}
\usepackage{graphicx}
\usepackage{mathtools}
\usepackage{amsmath,amssymb,amsfonts,amsthm,latexsym,stmaryrd,physics,mathrsfs}
\allowdisplaybreaks[4]
\usepackage{marginnote}
\usepackage{color}
\usepackage{bm}
\usepackage{float}
\usepackage{nicefrac}
\usepackage{microtype} 
\usepackage{booktabs}
\usepackage{verbatim}
\usepackage{setspace}
\usepackage[T1]{fontenc}
\usepackage[utf8]{inputenc}
\usepackage{stfloats}
\usepackage{diagbox}
\usepackage{dsfont}
\usepackage[colorinlistoftodos,prependcaption,textsize=tiny]{todonotes}
\usepackage{epstopdf}
\usepackage{latexsym}
\usepackage{xparse}
\usepackage[center]{subfigure}

\hypersetup{
colorlinks=true
,urlcolor=blue
,anchorcolor=blue
,citecolor=blue
,filecolor=blue
,linkcolor=blue
,menucolor=blue
,pagecolor=blue
,linktocpage=true
,pdfproducer=medialab
}
 
 \newcommand{\bq}{\begin{equation}}
 \newcommand{\eq}{\end{equation}}
 \newcommand{\bqn}{\begin{eqnarray}}
 \newcommand{\eqn}{\end{eqnarray}}
 \newcommand{\nb}{\nonumber}
     \newcommand{\lb}{\label}

\NewDocumentCommand{\evalat}{sO{\big}mm}{%
  \IfBooleanTF{#1}
   {\mleft. #3 \mright|_{#4}}
   {#3#2|_{#4}}%
}
\def\be{\begin{eqnarray}}
\def\ee{\end{eqnarray}}

\newcommand{\cc}{\mathcal C}

\newcommand{\calp}{\mathcal P}

\newcommand{\cs}{\mathcal S}

\newcommand{\lt}{\left}
\newcommand{\rt}{\right}

\def\R{\mathbb{R}}

\renewcommand{\b}{\beta}

\newcommand{\G}{\Gamma}

\newcommand{\sig}{\sigma}

\renewcommand{\l}{\lambda}

\renewcommand{\O}{\Omega}
\renewcommand{\t}{\tau}

\newcommand{\rmd}{\mathrm d}

\begin{document}
 
\title{Spherical symmetric gravitational collapse of a dust cloud: polymerized dynamics in reduced phase space}

\author{Kristina Giesel}
\email{kristina.giesel(AT)gravity.fau.de} 
\affiliation{Department Physik, Institut f\"ur Quantengravitation, Theoretische Physik III, Friedrich-Alexander Universit\"at Erlangen-N\"urnberg, Staudtstr. 7/B2, 91058 Erlangen, Germany}

\author{Muxin Han}
\email{hanm(AT)fau.edu}  
\affiliation{Department of Physics, Florida Atlantic University, 777 Glades Road, Boca Raton, FL 33431, USA}
\affiliation{Department Physik, Institut f\"ur Quantengravitation, Theoretische Physik III, Friedrich-Alexander Universit\"at Erlangen-N\"urnberg, Staudtstr. 7/B2, 91058 Erlangen, Germany}

\author{Bao-Fei Li}
\email{libaofei(AT)zjut.edu.cn}
\affiliation{Institute for Theoretical Physics $\&$ Cosmology, Zhejiang University of Technology, Hangzhou, 310032, China}
\affiliation{Department of Physics and Astronomy, Louisiana State University, Baton Rouge, LA 70803, USA}

\author{Hongguang Liu} 
\email{hongguang.liu(AT)gravity.fau.de}
\affiliation{Department Physik, Institut f\"ur Quantengravitation, Theoretische Physik III, Friedrich-Alexander Universit\"at Erlangen-N\"urnberg, Staudtstr. 7/B2, 91058 Erlangen, Germany}

\author{Parampreet Singh}
\email{psingh(AT)lsu.edu}
\affiliation{Department of Physics and Astronomy, Louisiana State University, Baton Rouge, LA 70803, USA}

\begin{abstract}
Based on the effective dynamics in the $\bar \mu$ scheme of the spherical symmetry reduced model in the reduced phase space formulation of loop quantum gravity (LQG), we investigate the gravitational collapse of a homogeneous dust cloud, with Gaussian dust serving as both the reference field and the source of the gravitational collapse. The effective dynamics from the considered model for a homogeneous dust cloud reduces precisely to the
effective dynamics of loop quantum cosmology (LQC) with extrinsic curvature based K-quantization, indicating that the LQC effective dynamics lives
as a subsector of the model presented here. In both the marginally bound and the bound cases of the collapse in effective dynamics, the singularity is resolved and replaced by a bounce. Though quantum geometric modification from spatial curvature is not directly included in the K-quantization it does affect the qualitative dynamics of the collapsing dust cloud in the sense that on the one hand for the marginally bound case, the dust cloud bounces once at fixed maximum energy density and on the other hand for the bound case, the dust cloud undergoes infinite cycles of contraction and expansion at energy densities dependent on the dust mass.  Finally, the mass threshold for the formation of a trapped surface in each case is found and the matching conditions between the interior collapsing spacetime and an effective exterior static solution are discussed.

\end{abstract}
\maketitle
\section{Introduction}
\label{sec:intro}

Non-perturbative quantum gravitational effects in LQG signal that the classical differential geometry of Einstein's gravity is replaced by a discrete quantum geometry at the Planck scale. The resulting quantum dynamics is expected to provide an upper bound on spacetime curvature and result in resolution of singularities. A rigorous demonstration of this happens in cosmological spacetimes in the context of loop quantum cosmology (LQC) where the classical big bang is replaced by a quantum big bounce \cite{Ashtekar:2006rx} and a generic resolution of strong curvature cosmological singularities occurs \cite{Singh:2009mz, Singh:2014fsy}. In recent years, investigations on similar lines have been carried out for black hole spacetimes (see e.g. \cite{Ashtekar:2005qt,Modesto:2005zm,Bohmer:2007wi,Chiou:2012pg,Gambini:2013hna,Corichi:2015xia,Dadhich:2015ora,Yonika:2017qgo, DAmbrosio:2020mut,Olmedo:2017lvt,Ashtekar:2018lag, Ashtekar:2018cay,Bojowald:2018xxu,Bodendorfer:2019cyv,Alesci:2019pbs,Assanioussi:2019twp,Benitez:2020szx, Gan:2020dkb, Kelly:2020lec,Gambini:2020nsf, Gambini:2020qhx, Husain:2022gwp, Li:2021snn, Gan:2022mle}, see also \cite{Ashtekar:2020ifw} for a review). The goal of most of these works is to capture the discreteness of quantum spacetime understood in LQG in an effective spacetime description which using high performance computing has proved to be reliable tool to understand quantum evolution using a set of quantum gravity modified dynamical equations \cite{Diener:2014mia, Diener:2017lde, Singh:2018rwa}. As in many above works, our work will be based on assuming validity of this effective spacetime description.

The models for describing black holes in the framework of LQG are mostly based on symmetry reduced models. Instead of the full quantum theory of gravity, these models reduce the degrees of freedom (DOFs) by implementing spherical symmetry at the classical level and quantize only the reduced set of DOFs satifying this symmetry. Such kind of symmetry reduced models have been successful in the study of quantum cosmology, see for instance \cite{Agullo:2016tjh} for a recent review.  The existing models of LQG black holes fall into two categories: the first category of models aim to quantize Schwarzschild black hole using isometry with Kantowski-Sachs vacuum cosmology   \cite{Ashtekar:2005qt,Modesto:2005zm,Bohmer:2007wi,Ashtekar:2010qz,Corichi:2015xia,Dadhich:2015ora,Yonika:2017qgo,Olmedo:2017lvt,Ashtekar:2018lag,Ashtekar:2018cay,Bodendorfer:2019cyv,Assanioussi:2019twp,Zhang:2020qxw,Zhang:2021wex}. These models quantize only a finite number of DOFs that result from requiring spherical symmetry as well as homogeneity. On the other hand the second category of models e.g. \cite{Bojowald:2005cb,Chiou:2012pg,Gambini:2013hna,BenAchour:2018khr,Alesci:2019pbs,Kelly:2020lec,Gambini:2020nsf,Han:2020uhb,Zhang:2021xoa} perform a symmetry reduction with respect to 
spherical symmetry only. These models give 1+1 dimensional field theories, which still contain infinitely many DOFs. Furthermore, both the black hole interior and exterior are treated in a unified manner in the second category. The model presented in this work belongs to the second category which is then reduced to explore the effects of quantum geometry on the collapse of a homogeneous dust cloud. In particular, our model is developed from the reduced phase space formulation of LQG. At the classical level and before the symmetry reduction, the gravitational field is coupled to Gaussian dust, which serves as reference fields to deparametrize gravity \cite{Kuchar:1990vy,Giesel:2012rb} and obtain the reduced phase space of the gravitational degrees of freedom in terms of elementary Dirac observables. This allows to formulate general relativity (GR) in a manifestly gauge invariant manner where the Hamiltonian and diffeomorphism constraints are solved classically and the dynamics of these Dirac observables is generated by a so called physical Hamiltonian that is non-vanishing in the physical sector of the theory. The symmetry reduction with respect to spherically  symmetry is then performed at the level of the reduced phase space and the Gauss constraint is solved with a gauge fixing leading to the constraint-free phase space $\calp$ of spherical symmetric physical DOFs. The quantization of the model is then applied to the reduced phase space and the algebra of Dirac observables directly. Therefore, the model is free of any complication arising from quantization of constraints. In fact, the quantum dynamics encoded in the physical Hamiltonian operator is manifestly unitary when passing to the quantum theory \cite{Zhang:2021xoa}. 

In this work, we apply the effective dynamics of the above physical Hamiltonian  ${\bf H}_\Delta$ to study the gravitational collapse for a homogeneous dust cloud. With dynamically coupling Gaussian dust to gravity we have on one hand the necessary reference fields in the system to construct the reduced phase space and on the other hand we can consider non-vacuum solutions such as the gravitational collapse involving dust as the source. The reduced phase space is derived in classical GR where for the Gaussian dust model the Dirac observable corresponding to the lapse function is unity and  the shift vector is zero. A question that arises when working at the effective level is how to carry over the form of the classical lapse and shift to the effective model. The strategy for this discussed recently in \cite{Giesel:2021rky} considers criteria that should be satisfied such that the effective versions for lapse and shift are  consistent with the effective dynamics. In general this means, that the effective lapse and shift are not just given by the polymerization of their classical counter parts \cite{Giesel:2021rky}. However, for the class of matter reference fields, including the Gaussian dust model, this is valid and it turns out that choosing the effective lapse and shift to be one and zero respectively is a consistent choice in this model.\footnote{The situation changes if one chooses geometric clocks as has for instance be done in \cite{Kelly:2020lec,corichi2016loop}. Then it depends whether the gauge fixing conditions involve variables that are or are not polymerized at the effective level. A procedure to obtain an effective lapse and shift consistent with the effective dynamics has been presented in \cite{Giesel:2021rky} and the analysis there shows that neither the model in \cite{Kelly:2020lec} nor the model in \cite{corichi2016loop} chooses a consistent lapse and shift if we assume that the gauge fixing conditions are just the polymerizations of their classical counter parts.}

Let us note that every quantization strategy is fraught with quantization ambiguities, and unlike the case of isotropic LQC, where mathematical and phenomenological considerations result in a unique quantization \cite{Corichi:2008zb, Corichi:2009pp} -- the so called $\bar \mu$ scheme or the improved dynamics \cite{Ashtekar:2006wn} and ruling out the old $\mu_o$ scheme in LQC. However, different quantization prescriptions have been put forth for black hole spacetimes. A priori there is no guarantee which scheme would be successful unless one probes the resulting physical implications in detail. Thus, a closer look at any of these prescriptions is necessary.  For the Schwarzschild black hole this task has been carefully carried out in the case of the interior spacetime where a recent study discussed inherent deficiencies of some of the schemes, including one based on quantizing Schwarzschild interior with a $\bar \mu$ scheme, and proposed a quantization prescription for Schwarschild interior \cite{Ashtekar:2018lag}. The situation for a gravitational collapse scenarios, which unlike Schwarzschild case, is a dynamical case as in cosmology is still to be settled. Results for homogeneous collapse in LQC setting rule out the $\mu_o$ scheme unless the Barbero-Immrizi parameter decreaes almost four times \cite{Li:2021snn}. It turns out that 
inside the dust cloud, the $\bar \mu$ effective dynamics improves the Oppenheimer-Snyder (OS) model by resolving the singularity with a non-singular bounce, where the curvature is Planckian. Although a large part of our discussion treats $k=0,\pm 1$ in general, we indeed focus on the improved OS models with $k=1$ (the bound case) and $k=0$ (the marginally bound case). The results on the marginally bound case with $k = 0$ are consistent with those obtained in \cite{Giesel:2021dug}, which uses a different symmetry reduction scheme. 
An important caveat of our analysis is that we only consider quantum geometric effects via polymerization of the extrinsic curvature and ignore the same for the intrinsic curvature. Basically we follow the so called  $K$-quantizations (where $K$ denotes extrinsic curvature) \cite{Singh:2013ava}. The bounce obtained here is time-reversal symmetric because of the simplifcation of ignoring quantum geometric effects to intrinsic curvature.

The dust cloud is assumed in our analysis to have a finite radius. The spacetime geometry outside the cloud is also needed in order to obtain a full description of the gravitational field. The effective spacetime outside the dust cloud should be governed by the same set of EOMs from ${\bf H}_\Delta$, and is matched to the Schwarzschild geometry far away from the dust cloud in \cite{Han:2020uhb}. Note that other matching conditions such as generalized Vaidya are also possible (see for eg. \cite{Giesel:2020raf}). In this paper we make the same assumption as in \cite{Han:2020uhb} for the exterior to investigate the matching condition on the dust shell between the effective spacetimes inside and outside the dust cloud, and we show that the matching condition is {\it{approximately}} satisfied except for the regime where quantum geometric effects become significant. The matching condition is not satisfied in the strong quantum regime due to quantum corrections to the Lemaitre–Tolman–Bondi (LTB) conditions, that restrict the spherical symmetric sector to the LTB solution, in the effective spacetime outside the dust cloud, while the LTB conditions are satisfied without correction for the effective spacetime inside the dust cloud.

There exists former results in the literature on the mass threshold of forming a horizon in the gravitational collapse with quantum geometric effects in different settings \cite{Bojowald:2005qw,Giesel:2021dug,Husain:2021ojz}. For a dust shell mass below a certain threshold, no horizon will form. With the model considered in this work, we confirm a mass threshold at the order of Planckian mass for both the marginally bound $k=0$ and the bound case $k=1$.  In our results, for a dust shell mass larger than the threshold, as the dust profile falls inward, a pair of apparent horizon forms before the bounce. There is no shock wave after the bounce, in contrast to the result in \cite{Husain:2021ojz}. For an observer that sits inside or on the dust shell, the trapped region will be reached first near the Schwarzschild radius. With the collapse continuing in the trapped region, an inner apparent horizon will be reached at the Planck curvature scale before the bounce, where the space-time region becomes untrapped. After that, matter bounces outward symmetrically. The observer will enter the anti-trapped white-hole region and finally moves out after crossing the white-hole horizon. The result here in the $k=0$ case agrees with the earlier result in \cite{Giesel:2021dug}.

The paper is organized as follows. After the introduction in section \ref{sec:intro} we review in section \ref{sec:review} the relational dynamics of the Gaussian dust model (subsection \ref{sec:RPSQ}) as well as its symmetry reduction to spherical symmetry at the classical level (subsection \ref{sec:SphSymmRed}). In addition in subsection \ref{sec:RevmubareffectDyn} we briefly review the effective dynamics in the $\bar \mu$-scheme following from the models in \cite{Han:2020chr,zhang2020loop}. Section \ref{sec:homogeneous-collapse} introduces the collapsing dust model of this work by further imposing two LTB conditions that restrict the spherical symmetric effective dynamics to the LTB sector. We start in subsection \ref{sec:RevLTBdustshell} with a brief summary on the LTB dust shell model from \cite{Giesel:2021dug} for which the symmetry reduction has been performed at the classical level along the lines of \cite{Kiefer:2019csi} before effective techniques have been applied. In subsection \ref{sec:OSdustmodel} we analyse the case of a homogeneous dust cloud with analytical methods. The model can be understood as an Oppenheimer-Snyder dust collapsing model, for the marginally bound case ($k=0$) and bound case with $k=1$. Further, we discuss in subsection \ref{sec:hamiltonian-and-conserved-quantity} the effective Hamiltonian as well as the resulting modified Friedmann-like equations for the $k=0$ and $k=1$ model and we obtain  a qualitatively different behavior of the two cases. Subsection \ref{sec:null-expansion-trapped-region-inner-horizon} analyses the formation of trapped surfaces in the model and derives the the resulting value for the threshold mass that agrees for the $k=0$ with one found in \cite{Giesel:2021dug}. Possible matching conditions for gluing the interior spacetime to an exterior stationary spacetime are given in subsection \ref{sec:MatchingCond}. In section \ref{sec:numerics} we present some numerical results for the model.

\section{Reduced phase space quantisation, spherical symmetry reduction, and effective dynamics}
\label{sec:review}
\renewcommand{\theequation}{2.\arabic{equation}}\setcounter{equation}{0}

In this section, we give a brief review on the reduced phase space formulation for gravity coupled to Gaussian dust and the sperical symmetry reduction. We also review briefly the $\bar{\mu}$-scheme effective dynamics of spherical symmetric LQG developed recently in \cite{Han:2020uhb}.

\subsection{Reduced phase space of the Gaussian dust model}
\label{sec:RPSQ}

Usage of reference fields  in GR in order to access the physical phase space and after quantization the physical Hilbert space respectively has been studied earlier \cite{Kuchar:1990vy,Kuchar:1991pq,Brown:1994py,Bicak:1997bx,Giesel:2007wn,Domagala:2010bm,Husain:2011tk,Giesel:2016gxq}. In particular, a classification of the existing scalar field reference models in the context of LQG can be found in \cite{Giesel:2012rb}. The individual models differ by the number and kind of reference fields that one couples dynamically to GR. In full GR in one type of models one has four additional reference fields that can be used to reduce the Hamiltonian as well as the spatial diffeomorphism constraint at the classical level. The other kind of models involve only one reference field that is typically used to reduce the Hamiltonian constraint, whereas the spatial diffeomorphism constraint is then solved in the quantum theory.  In this work we will focus on models that allow to reduce the Hamiltonian as well as the diffeomorphism constraint and these models have in common that one obtains  a system with second class constraints. In these models one couples eight or seven additional fields to gravity and after reduction with respect to the second class constraints one ends up with a first class systems which has four additional fields to the geometric degrees of freedom. In case one takes GR in terms of Ashtekar-Barbero variables as a starting point as for instance has been done in \cite{Giesel:2006um,Domagala:2010bm,Giesel:2009jp} the models involve an additional Gauss constraint that is solved via Dirac quantization in the quantum theory by working with gauge invariant spin network functions. Another alternative is to gauge-fix the Gauss constraint already at the classical level as it is often done in symmetry reduced models. The different dust models available in the literature for full GR and corresponding quantum gravity models \cite{Brown:1994py,Kuchar:1990vy,Giesel:2006um,Husain:2011tk,Giesel:2012rb} carry different features at the level of full GR such as for instance a different number of coupled dust fields and a different form of the resulting physical Hamiltonian. Once symmetry reduced to FLRW spacetimes most of the distinctive properties are lost due to the simplicity of the models and in particular because the spatial diffeomorphism constraint vanish trivially in these models, see for instance \cite{Giesel:2020raf} where different reference matter models have been analysed in the cosmological context involving in addition to dust also Klein-Gordon scalar fields as reference matter. This is no longer the case for spherically symmetric models where more distinguishable properties of the dust models are present. In the current work we will focus on the Gaussian dust model that was introduced in the seminal article \cite{Kuchar:1990vy} and see \cite{Giesel:2006um} for the corresponding quantum model using a loop quantization. If we choose dust as reference fields in spherically symmetric models, we work in the framework of LTB models. The main motivation for focusing on the Gaussian dust model here is that its physical Hamiltonian has a simpler form than in the Brown-Kucha\v{r} model. 

The Gaussian dust model considers the following total action
\begin{eqnarray*}
S_{\rm tot}& =& S_{\rm geo}+S^{\rm G}_{\rm dust}
\end{eqnarray*}
with the geometric part described by the Einstein-Hilbert action
\begin{equation}
S_{\text {geo }}=\frac{1}{2 \kappa} \int_{\mathcal{M}} \mathrm{d}^{4} x \sqrt{-g} R^{(4)}
\end{equation}
whereas the dust dynamics is encoded in the following action
\begin{equation}
S_{\mathrm{dust}}^{\mathrm{G}}=-\int \mathrm{d}^{3} x \sqrt{-g}\left(\frac{\rho}{2}\left[g^{\mu \nu} T_{, \mu} T_{, \nu}+1\right]+g^{\mu \nu} T_{, \mu} W_{j} S_{\nu}^{j}\right),
\end{equation}
here $\kappa=8\pi G$ with $G$ being Newton's constant and $\rho,T,W_j,S^j$ with $j=1,2,3$ denote eight dynamically coupled scalar fields describing the Gaussian dust model. After an ADM decomposition, the kinematical phase space consists of the ADM variables of $(q_{ab},p^{ab},N,P_N,N^a,P_a)$ in the gravitational sector, where $q_{ab}$ denotes the ADM metric, $N$ the lapse function and $N^a$ the shift vector. The kinematical degrees of freedom in the dust sector are given by $(\rho,P_\rho,T,P_T, W_j, P_{W_j},S^j,P_j)$ so that in total the model involves 36 phase space degrees of freedom at the kinematical level. As far as the constraints of the Gaussian dust model are concerned it is  a second class constrained system and as shown in \cite{Kuchar:1990vy,Giesel:2006um} after the reduction of the second class constraints the independent variables  are given by the set $(q_{ab},p^{ab},T,P_T,S^j,P_j)$ and this partially reduced system becomes first class for more details as well es the explicit form of the constraints compatible with our notation see \cite{Giesel:2012rb}. We take this as a starting point for the work in this article and consider the usual extension of the gravitational phase space from ADM to Ashtekar-Barbero variables denotes by $(A_a^j,E^a_j)$ which are a SU(2)-connection and a densitised triad respectively building a canonical pair with the non-vanishing Poisson brackets
\begin{equation*}
\{A_a^j(x),E^b_k(y)\} =\frac{\kappa\beta}{2}\delta^j_k\delta_a^b\delta^{(3)}(x,y),   \quad \kappa=16\pi G. 
\end{equation*}
Here $A^j_a:=\Gamma^a_j+\beta K^j_a$ where $\Gamma^j_a$ is the spin connection, $\beta$ the Immmirzi parameter and $K^j_a$ is related to the extrinsic curvature via $K^j_a=e_j^b K_{ab}$ with $e^j_b$ denoting the usual co-triads. 
The remaining first class constraints are the SU(2) Gauss constraint $G_j$, the spatial diffeomorphism constraint $c_a^{\rm tot}$ and the Hamiltonian constraint $c^{\rm tot}$. The total Hamiltonian constraint consisting of the geometric and dust contributions denoted by $c$ and $c^{\rm dust}$ respectively reads
\begin{eqnarray}
c^{\rm tot}&=& c + c^{\rm dust} \\
c&=&\frac{1}{2\kappa} \frac{\varepsilon_{j}^{mn} E_{m}^{a} E_{n}^{b}}{\sqrt{|\operatorname{det}(E^a_j)}|}\left(F_{ab }^{j}-\left(1+\beta^{2}\right) \varepsilon^{jk\ell} K_{a}^{k} K_{b}^{m}\right)  \nonumber\\
c^{\rm dust}&=&P \sqrt{1+\frac{E^a_mE^b_n\delta^{mn}T_{, a} T_{, b}}{\det(E^a_j)} }+\frac{E^a_mE^b_n\delta^{mn}}{\det(E^a_j)}\frac{T_{, a}P_{j} S_{, b}^{j}}{\sqrt{1+\frac{E^a_mE^b_n\delta^{mn}T_{, a} T_{, b}}{\det(E^a_j)} }}, \nonumber
\end{eqnarray}
where 
\begin{equation*}
  F_{ab}^{j}=\partial_{a} A_{b}^{j}-\partial_{b} A_{a}^{j}+\epsilon^{j}{}_{k\ell} A_{a}^{k} A_{b}^{\ell}  
\end{equation*}
and $K_{a}^{j}=\frac{1}{\beta}\left(A_{a}^{j}-\Gamma_{a}^{j}\right)$ is considered as a function of $(A^j_a,E^a_j)$. The total spatial diffeomorphism constraint is given by    
\begin{eqnarray}
c^{\rm tot}_a&=& c_a + c^{\rm dust}_a \quad {\rm with}\quad 
c_a=\frac{1}{\kappa\beta}F^j_{ab}E^a_j \quad 
c^{\rm dust}_a=\frac{1}{\kappa\beta}\left(PT_{,a}+P_jS^j_{,a}\right) 
\end{eqnarray}
and the total Gauss constraint takes the form
\begin{equation*}
G_{j}=\frac{1}{\kappa\beta}\left(\partial_{a} E_{j}^{a}+\epsilon_{\ j k}^{\ell} A_{a}^{k} E_{\ell}^{a}\right).
\end{equation*}
As presented in \cite{Kuchar:1990vy,Giesel:2006um} the Hamiltonian as well as the diffeomorphism constraint can be solved for the dust momenta $P$ and $P_j$ respectively allowing to work with an equivalent set of these constraints being all linear in the dust momenta. As a consequence, the dust fields $T$ and $S^j$ are canonically conjugate to the total Hamiltonian and diffeomorphism constraint respectively and thus provide good candidates for reference fields for these constraints. Applying the observable map from \cite{Vytheeswaran:1994np,Dittrich:2004cb,Dittrich:2005kc} in the framework of the relational formalism \cite{Rovelli:1990ph,Rovelli:1990pi,Rovelli:2001bz}, as presented in \cite{Giesel:2006um}, we can construct Dirac observables corresponding to the canonical pair $(A_a^j,E^a_j)$. Following the notation from \cite{Giesel:2006um,Giesel:2012rb} we denote these Dirac observables by $(A_j^J,E^j_J)$. Here $j$ is the index labelling coordinates on the dust manifold ${\cal S}$ with coordinates $\sigma^j$ being those values the dust field $S^j$ take under the observable map and  $J$ is a su(2)-index both running from 1 to 3. As shown in \cite{Giesel:2012rb} the algebra of the Dirac observables has a standard canonical form with
\begin{equation*}
\{A_j^J(\sigma,\tau), E^k_K(\sigma',\tau)\} =\frac{\kappa\beta}{2}\delta_j^k\delta^J_K\delta^{(3)}(\sigma,\sigma'), 
\end{equation*}
where $\tau$ denotes  physical time related to the reference field $T$ and all remaining Poisson bracket vanish. $(A_j^J,E^j_J)$ are the elementary variables of the reduced phase space. The dynamics of these Dirac observables is generated by a so called physical Hamiltonian, that is itself a Dirac observable and not vanishing on the constraint hypersurface. For the Gaussian dust model it has the following form \cite{Giesel:2012rb}
\begin{equation*}
{\bf H}^{\rm G}_{\rm phys} = \int\limits_{\cal S}d^3\sigma \mathcal{C}(\sigma)    
\end{equation*}
where $\mathcal{C}(\sigma)$ denotes the Dirac observable of $c(x)$ given by
\begin{equation*}
\mathcal{C}=\frac{1}{2\kappa} \frac{\varepsilon_{J}^{MN} E_{M}^{j} E_{N}^{k}}{\sqrt{|\operatorname{det}(E^j_J)}|}\left(F_{jk }^{J}-\left(1+\beta^{2}\right) \varepsilon^{J K L} K_{j}^{K} K_{k}^{M}\right)    .
\end{equation*}
Here $F_{jk}^{J}$ is the curvature associated with the connection $A_{j}^{J}$
$$
F_{jk}^{J}=\partial_{j} A_{k}^{J}-\partial_{k} A_{j}^{J}+\epsilon_{\ KL}^{J} A_{j}^{K} A_{k}^{L}
$$
and $K_{j}^{J}=\frac{1}{\beta}\left(A_{j}^{J}-\Gamma_{j}^{J}\right)$ is considered as a function of the elementary Dirac observables $(A^J_j,E^j_J)$. To obtain the Dirac observable ${\cal C}$ we took advantage of the fact that for the observable map we have $\mathcal{C}=c(A^J_j,E^j_J)$ and likewise for any other function on the reduced phase space as for instance $c_a$ and $G_j$, a derivation of the detailed properties of the observable can be found in \cite{Dittrich:2004cb,Dittrich:2005kc,Thiemann:2004wk}.
This reduced phase space as well as the dynamics encoded in ${\bf H}^{\rm G}_{\rm phys}$ at the classical level will be our starting point for considering a spherically symmetric symmetry reduction in the next subsection.
\subsection{Spherical symmetry reduction}
\label{sec:SphSymmRed}

In this work, we focus only on the sector of spherical symmetrical degrees of freedom in the reduced phase space, and restrict the LQG dynamics to the spherical symmetrical degrees of freedom. Our scheme is similar to e.g. \cite{Ashtekar:2005qt,Bojowald:2005cb,Chiou:2012pg,Gambini:2013hna,Zhang:2021xoa,Han:2020uhb}. For obtaining the spherically symmetric midisuperspace we assume the dust space $\cs\simeq \R\times S^2$ and define the spherical coordinate ${\sig}=(x,\theta,\phi)$. We restrict the reduced phase space to the phase space $\Gamma_{\rm red}$ of spherical symmetric field configurations. In spherically symmetric spacetimes, one only considers $(A^I_j,E^j_I)$ that are invariant under rotations up to gauge transformations. The general forms are given by 
\begin{equation}
\label{eq:AEexpress}
\begin{aligned}
A^I_j\tau_I\dd \sigma^a=&A_1(x)\tau_1\dd x+\frac{1}{\sqrt{2}}( A_2(x)\tau_2+ A_3(x)\tau_3)\dd\theta+\frac{1}{\sqrt{2}}( A_2(x)\tau_3-A_3(x)\tau_2)\sin(\theta)\dd\varphi+\cos(\theta)\tau_1\dd\varphi,\\
E_I^j\tau^I\frac{\partial}{\partial\sigma^a}=&E^1(x)\sin(\theta)\tau_1\partial_x+\frac{1}{\sqrt{2}}( E^2(x)\tau_2+ E^3(x)\tau_3)\sin(\theta)\partial_\theta+\frac{1}{\sqrt{2}}( E^2\tau_3-E^3\tau_2)\partial_\varphi,
\end{aligned}
\end{equation}
where $\tau_I=-\frac{i}{2}\sigma_I$ with $\sigma_I$ denoting Pauli matrices. We denote by $\Gamma_{\rm red}$ the reduced phase space of the spherically symmetric $(A^I_j,E^j_I)$. The symplectic form $\Omega$ on $\Gamma_{\rm red}$ reads
\begin{equation}\label{eq:symplecticform}
\begin{aligned}
\Omega(\delta_1,\delta_2)=&-\frac{2}{\kappa\beta}\int\dd^3\sigma \delta_1 A_j^I(\sigma)\wedge\delta_2 E^j_I(\sigma)\\
=&-\frac{1}{2G\beta}\int\left(\delta_1A_1(x)\wedge \delta_2 E^1(x)+\delta_1 A_2(x)\wedge\delta_2 E^2(x)+\delta_1A_3(x)\wedge\delta_2 E^3(x) \right)\dd x,
\end{aligned}
\end{equation}
where $\delta_1$ and $\delta_2$ are differentials on $\Gamma_{\rm red}$. The Poisson bracket from $\O$ implies $\{A_j(x),E^k(x')\}=2G\beta\delta(x,x')\delta_j^k$, with $j,k=1,2,3$. The symmetry-reduced theory is an (1+1)-dimensional field theory with infinite-dimensional $\G_{\rm red}$.

We still need to impose the Gauss constraint to $\Gamma_{\rm red}$. Eqs.\eqref{eq:AEexpress} reduce the Gauss constraint to only one constraint:
\begin{equation}\label{eq:gaussian}
G[\lambda]=4\pi\int \rmd x\,\lambda(x)\left[A_2(x)E^3(x)-A_3(x)E^2(x)+\partial_xE^1(x)\right],
\end{equation}
while other two components become trivial. Correspondingly, the SU(2) gauge group is reduced to U(1). Under the gauge transformation generated by $G[\l]$, $A_1$ and $E^1$ transform as a U(1) gauge field and electric field, while $A_2+i A_3$ and $E^2+iE^3$ transform as U(1) scalar fields: 
\be
A_1(x)&\to &A_1(x)-\kappa\beta\partial_x\lambda^1(x),\qquad \qquad E^1(x)\to E^1(x),\label{U1gaugef} \\
A_2(x)+iA_3(x)&\to&e^{i\kappa\beta\lambda^1(x)}(A_2(x)+i A_3(x)),
E^2(x)+iE^3(x)\to e^{i\kappa\beta\lambda^1(x)}(E^2(x)+i E^3(x)).
\ee
One can always gauge transform $(A_I,E^I)\in\Gamma_{\rm red}$ to make $E^3$ vanish. Thus we introduce the following gauge fixing condition 
\begin{equation}
E^3(x)=0.\label{gaugefix}
\end{equation}
Then we solve the Gauss constraint \eqref{eq:gaussian} for $A_3(x)$
\begin{equation}
A_3(x)=\frac{\partial_xE^1(x)}{E^2(x)}.\label{A3}
\end{equation}
Eqs.\eqref{gaugefix} and \eqref{A3} remove $(A_3,E^3)$ from the canonical pairs. Following \cite{Han:2020uhb,Zhang:2021xoa,Gambini:2013hna}, we introduce the following variables
\begin{equation}
\begin{aligned}
K_x(x):=\frac{1}{2\beta}A_1(x),\quad K_\varphi(x):=\frac{1}{\sqrt{2}\beta}A_2(x),\quad  E^x(x)=E^1(x),\quad E^\varphi(x)=\frac{1}{\sqrt{2}}E^2(x).
\end{aligned}
\end{equation}
The gauge-fixed reduced phase space, denoted by $\calp$, consists of canonical pairs $(K_x(x),E^x(x))$ and $(K_\varphi(x),E^\varphi(x))$ with the Poisson brackets
\begin{equation}
\label{Poisson1}
\{K_j(x),E^k(x')\}=G\delta^k_j\delta(x,x'),\quad j,k=x,\varphi.
\end{equation}
In terms of these variables, the metric is given by
\bq
\label{metric1}
\rmd s^2=-\rmd t^2+\frac{(E^\varphi)^2}{|E^x|}\rmd x^2+|E^x|\rmd\Omega^2,
\eq
where the angular part $\rmd\Omega^2=\rmd\theta^2+\sin^2\theta \rmd\varphi^2$.

The classical physical Hamiltonian ${\bf H}^{\rm G}_{\rm phys}$ reduced to the (gauge-fixed) spherical symmetrical sector $\calp$ gives 
\be
\mathbf{H}_{0} &=&\int \rmd x\, \mathcal{C}(x) + \text{boundary term},\label{H0CBDY}\\
 \\
\mathcal{C}(x) &=&\frac{4 \pi}{\kappa} \frac{\operatorname{sgn}\left(E^{\varphi}\right)}{\sqrt{\left|E^{x}\right|}}\left(-\frac{2 E^{x} E^{x \prime} E^{\varphi \prime}}{E^{\varphi 2}}+\frac{4 E^{x} E^{x \prime \prime}+E^{x \prime 2}}{2 E^{\varphi}}-8 E^{x} K_{x} K_{\varphi}-2 E^{\varphi}\left[K_{\varphi}^{2}+1\right]\right).
\ee
where ${E^x}'=\partial_x E^x$. For completeness we also mention the total diffeomorphism constraint that will be needed for the later discussion in section \ref{sec:homogeneous-collapse} and is given by 
\begin{equation}
\label{eq:ClassDiffeo}
\mathcal{C}^{\rm tot}_{x}(x)=P_x(x)+\mathcal{C}_x(x)=P_x(x)+E^{\varphi}(x) K_{\varphi}^{\prime}(x)-K_{x}(x) E^{x \prime}(x) \approx 0   
\end{equation}
where $P_x$ denotes the dust momentum conjugate to the reference field $S^x$. Here $\cc_x(x)$ for all $x$ are infinitely many conserved charges satisfying $\{\cc_x(x),{\bf H}_0\}=0$.

A boundary term in terms of Ashtekar-Barbero variables in the case of asymptotically flat spacetimes has been discussed in the literature \cite{thiemann1995generalized,Corichi:2013zza,Campiglia:2014yja}. In the following we briefly discuss how the boundary term for spherically symmetric spacetimes given in \eqref{H0CBDY} can be obtained: When deriving EOMs from ${\bf H}_0$, the variation $\delta\int\rmd x\, \cc(x)$ and the integration by parts result in the following boundary terms
\begin{eqnarray}
\frac{8 \pi  E^x{} \delta E^x{}'{}}{\kappa  \sqrt{\left| E^x{}\right| } \left| E^\varphi{}\right| }-\frac{8 \pi  E^x{} \delta E^\varphi {} \left| E^\varphi{}\right|  E^x{}'{}}{\kappa  E^\varphi{}^3 \sqrt{\left| E^x{}\right| }}.\label{bdyterm}
\end{eqnarray}
These boundary variations should be cancelled by the variation of the boundary term in \eqref{H0CBDY} with certain boundary condition, in order to have the well-defined variation. We are interested in the following boundary conditions:

\begin{itemize}

\item The LTB conditions (see \eqref{mb LTB conditions} and set $P_x=0$) restricts the spherical symmetric spacetimes to LTB spacetimes, and here the boundary condition involves one of the LTB condition $E^x{}'=2f(x)E^\varphi$ for a given function $f(x)$, see Section \ref{sec:homogeneous-collapse} for details. Since we are going to study the LTB dust shell model, we are interested in this LTB boundary condition. The boundary can be of finite distance or at infinity. The LTB condition implies $[\delta E^x{}'-2f(x)\delta E^\varphi]_{bdy}=0$. In this case, the two terms in \eqref{bdyterm} cancel each other in any variation $\delta\int\rmd x\, \cc(x)$ satisfying the LTB boundary condition. So we can set the boundary term to be zero in \eqref{H0CBDY}.

\item When we study the dynamics of spherical symmetric black hole, we consider $E^x,E^\varphi$ to behave asymptotically as the Schwarzschild geometry in the Lema\^{\i}tre coordinates as $x\to\infty$ \footnote{The Schwarzschild spacetime in the Lema\^{\i}tre coordinates $(t,x,\theta,\varphi)$ is given by \eqref{metric1} with $E^x= \left(\frac{3}{2} \sqrt{R_s}\, (x-t)\right)^{4 / 3},\  E^\varphi=  \sqrt{{R_s}} \lt(\frac{3}{2}{\sqrt{{R_s}}\, (x-t)}\rt)^{1/3}$. }:
\begin{eqnarray}
E^x\sim \left(\frac{3}{2} \sqrt{R_s}\, x\right)^{4 / 3},\quad E^\varphi\sim  \sqrt{{R_s}} \lt(\frac{3}{2}{\sqrt{{R_s}}\, x}\rt)^{1/3}, \label{bdyschw1}
\end{eqnarray}
where $R_s$ is the Schwarzschild radius. The LTB boundary condition  $E^x{}'=2E^\varphi$ is satisfied asymptotically. So we have again zero boundary term in \eqref{H0CBDY} at $x\to\infty$ for the asymptotic Schwarzschild boundary condition.

\item Alternatively, we may consider an infrared cut-off of the dust space at boundary (${bdy})=\{x=L\gg1\}$ and impose the Dirichlet boundary condition $\delta E^x|_{bdy}=0$. In this case, we have to add the following boundary term to the physical Hamiltonian
\begin{eqnarray}{\bf H}_0=\int_{-\infty}^\infty\rmd x\, \cc(x)+{\bf H}_{bdy},\quad{\bf H}_{bdy}=-\frac{8\pi}{\kappa}\lt(\frac{\sqrt{E^{x}}E^{x}{}^\prime{}}{E^{\varphi}}-2f(x)\sqrt{E^x}\rt)\Bigg|_{bdy},
\end{eqnarray}
for any function $f(x)$. $\delta_{E^\varphi(x)}{\bf H}_{bdy}$ cancels the boundary terms from $\delta_{E^\varphi(x)}\int_{-\infty}^\infty\rmd x\, \cc_\Delta(x)$, while $\delta_{E^x(x)}{\bf H}_{bdy}$ cancels the boundary term from $\delta_{E^x(x)}\int_{-\infty}^\infty\rmd x\, \cc_\Delta(x)$ up to a term proportional to $\delta E^x$ which vanishes by the Dirichlet boundary condition.

\item The Neumann boundary condition $E^x{}'|_{bdy}= 0,\ \delta E^x{}' |_{bdy}=0 $ is interesting in the discussion in \cite{Han:2020uhb} as $x\to-\infty $ as a part of the Nariai limit. Both terms in \eqref{bdyterm} vanish by this boundary condition, so no boundary term is needed. 

\end{itemize}

\subsection{Review of the effective dynamics in the $\bar \mu$ scheme}\label{review scheme}
\label{sec:RevmubareffectDyn}

For studying the LQG corrections to the spherical symmetrical spacetimes, the effective dynamics improved from the classical dynamics by ${\bf H}_0$ is developed in \cite{Han:2020uhb,Zhang:2021xoa}, where a $\bar{\mu}$-scheme improved Hamiltonian ${\bf H}_{\Delta}$ is defined by implementing the LQG holonomy corrections to ${\bf H}_0$. In the following, we briefly discuss a few key points in constructing ${\bf H}_{\Delta}$ and readers can refer to \cite{Han:2020uhb,Zhang:2021xoa} for details.

\begin{itemize}
    \item \emph{Spherical symmetry reduction and gauge fixing before quantization}: The starting point of the construction is the spherical symmetric form \eqref{eq:AEexpress} of $(A_j^I,E^j_I)$ on the dust space. The classical phase space of the full theory is reduced to a subspace $\G_{\rm red}$ of spherical symmetrical fields. $\G_{\rm red}$ is further reduced to $\calp$ by the gauge fixing $E^3=0$ and solving the Gauss constraint. In particular, $E^j_I$ is diagonal in this gauge. All the further development, including the improved Hamiltonian and quantization, are based on $\calp$.

    \item \emph{The U(1) holonomy of $A_1$, and the point holonomy of $A_2$}: The triad variables $E^j_I$ reduces to $E^1,E^2$ on $\calp$. We choose $A_1$ and $A_2$ to be their conjugate variables, and define $e^{i\l \int_e\rmd x A_1}$ and $e^{i\mu A_2}$ as the basic variables in the quantization and the regularization of the Hamiltonian \cite{Zhang:2021xoa}. The choice of the U(1) holonomy $e^{i\l \int_e\rmd x A_1}$ is natural since $A_1$ transforms as the U(1) gauge field, see \eqref{U1gaugef}. However, the component of $A_j^I$ perpendicular to $x$-direction can only give the holonomy supported at a point in the space of $x$. For simplicity, we choose the point holonomy $e^{i\mu A_2}$ of $A_2$ as the other basic variable\footnote{Although we focus on this choice in the present paper, we would like to mention that the alternative choice may be to use $e^{i\mu [A_2+f(E)]}$ for a certain nontrivial function $f(E)$ of $E^j$. The alternative choice may lead to a more complicated expression of ${\bf H}_\Delta$.  }

\item \emph{The $\bar{\mu}$-scheme regularization of the Hamiltonian with holonomies of fixed lengths. }: When constructing ${\bf H}_{\Delta}$, the U(1) holonomy and point holonomies are represented as belonging to U(1) subgroups in SU(2): $e^{\int_e \rmd x \bar{\l} A_1\t_1}$ and $e^{\bar{\mu} A_2\t_2},e^{\bar{\mu} A_2\t_3}$. Here $\t_1,\t_2,\t_3$ generate the U(1) subgroups of SU(2). These SU(2) holonomies are used for regularizing the SU(2) curvature: $ F\simeq\frac{1}{\Delta}[h_\Delta(\Box)-1]$. $h_\Delta(\Box)$ is the SU(2) loop holonomy around the plaquette $\Box$  whose area is fixed to be $\Delta$. We express $h_\Delta(\Box)$ in terms of holonomies along edges of fixed length $\sqrt{\Delta}$ \cite{Han:2020uhb}
\be
&&h_\Delta^1=e^{ \int_e \rmd x\bar{\l} A_1\t_1}\simeq e^{\frac{\sqrt{\Delta}\sqrt{|E^x|}}{{|E^\varphi|}}\,2\b K_x\t_1},\quad h_\Delta^2=e^{\bar{\mu} A_2\t_2}=e^{\frac{\sqrt{\Delta}}{\sqrt{|E^x|}}\,\b K_\varphi\t_2},\quad \\
&&h_\Delta^3=e^{\bar{\mu} A_2\t_3}=e^{\frac{\sqrt{\Delta}}{\sqrt{|E^x|}}\,\b K_\varphi\t_3},\quad h_\Delta(\Box_{jk})= h_\Delta^jh_\Delta^k(h_\Delta^j)^{-1}(h_\Delta^k)^{-1},\quad j,k=1,2,3.
\ee
We can regularize the $K$-dependent terms in ${\bf H}_0$ in terms of these holonomies and construct 
\be
{\bf H}_{\Delta}&=&\frac{2}{\b^2\kappa\Delta}\int\rmd^3x\sum_{j,k}e(\Box_{jk})\mathrm{Tr}\lt(h_\Delta(\Box_{jk})\frac{[E^j,E^k]}{\sqrt{\det(q)}}\rt)+\text{terms independent of}\ K.
\ee
\end{itemize}

As the result from the above discussion, we obtain the following expression of the $\bar{\mu}$-scheme improved Hamiltonian ${\bf H}_{\Delta}$ defined on $\calp$ \cite{Han:2020uhb} 
\be
\lb{Hamiltonian}
\mathbf{H}_\Delta&=&\int \dd x\, \cc_\Delta(x)+\text{boundary term},\\
\lb{Hamiltonian density}
\cc_\Delta(x)&=&\frac{1}{4G}\frac{\text{sgn}(E^{\varphi}{})}{\sqrt{\left|E^{x}{}\right|}}\Bigg(-\frac{2E^{x}{}E^{x}{}^\prime{}E^{\varphi}{}^\prime{}}{E^{\varphi}{}^{2}}+\frac{4E^{x}{}E^{x}{}^{\prime\prime}{}+E^{x}{}^\prime{}^{2}}{2E^{\varphi}{}} \nonumber\\
&&-\frac{4E^{x}E^\varphi}{\b^2\Delta}\sin\lt[\frac{\sqrt{\Delta}\sqrt{|E^x|}}{{E^\varphi}}2\b K_x(x)\rt]\sin\lt[\frac{\sqrt{\Delta}}{\sqrt{|E^x|}}\b K_\varphi(x)\rt]\nonumber\\
&&-\frac{2E^{\varphi}{|E^x|}}{\b^2{\Delta}}\sin^2\lt[\frac{\sqrt{\Delta}}{\sqrt{|E^x|}}\b K_\varphi(x)\rt]-2E^{\varphi}\Bigg). \label{HDelta2}
\ee
Effectively, $\mathbf{H}_\Delta$ improves the classical spherical symmetric Hamiltonian ${\bf H}_0$ by implementing the holonomy corrections 
\be
K_\varphi(x)\to \frac{\sqrt{|E^x|}}{\b\sqrt{\Delta}}\sin\lt[\frac{\sqrt{\Delta}}{\sqrt{|E^x|}}\b K_\varphi(x)\rt],
K_x(x)\to \frac{E^\varphi}{2\b\sqrt{\Delta |E^x|}}\sin\lt[\frac{\sqrt{\Delta |E^x|}}{{E^\varphi}}2\b K_x(x)\rt],\label{mubarKx}
\ee
where the deformation parameter $\Delta$ is assumed to be the same order of magnitude as the minimal area gap in LQG. Clearly as $\Delta\to0$, \eqref{mubarKx} reduces to $K_x,K_\varphi$. It is straightforward to check $\{\cc_x(x),{\bf H}_\Delta\}=0$, thus the conservation of $\cc_x(x)$ carries over to the improved dynamics. Here the conserved $\mathcal{C}_x(x)=E^{\varphi}(x) K_{\varphi}^{\prime}(x)-K_{x}(x) E^{x \prime}(x)$ does not contain any holonomy correction. The discussion of the boundary term in ${\bf H}_\Delta$ is exactly the same as the discussion for ${\bf H}_0$, since the boundary terms from $\delta \int \rmd x\, \cc_{\Delta}(x)$ is the same as \eqref{H0CBDY}.

The effective dynamics of the spherical symmetric gravity-dust system is given by the Hamiltonian equations from ${\bf H}_\Delta$ \cite{Han:2020uhb}: 

\be
\partial_t K_x &=& -\frac{\partial_x E^x \partial_x E^\varphi}{4 \sqrt{E^x{}} E^\varphi{}^2}-\frac{\left(\partial_x E^x\right)^2}{16 E^x{}^{3/2} E^\varphi{}}+\frac{\partial^2_x E^x}{4
   \sqrt{E^x{}} E^\varphi{}}+\frac{E^\varphi{}}{4 E^x{}^{3/2}}\nonumber\\
&-&\frac{E^\varphi{} \sin
   \left(\frac{\beta  \sqrt{\Delta } K_\varphi{}}{\sqrt{E^x{}}}\right) \sin \left(\frac{2 \beta 
   \sqrt{\Delta } \sqrt{E^x{}} K_x{}}{E^\varphi{}}\right)}{2 \beta ^2 \Delta 
   \sqrt{E^x{}}}-\frac{K_x{} \sin \left(\frac{\beta  \sqrt{\Delta } K_\varphi{}}{\sqrt{E^x{}}}\right) \cos \left(\frac{2 \beta  \sqrt{\Delta } \sqrt{E^x{}} K_x{}}{E^\varphi{}}\right)}{\beta  \sqrt{\Delta }}\nonumber\\
&+&\frac{E^\varphi{} K_\varphi{} \cos \left(\frac{\beta  \sqrt{\Delta }
   K_\varphi{}}{\sqrt{E^x{}}}\right) \sin \left(\frac{2 \beta  \sqrt{\Delta } \sqrt{E^x{}}
   K_x{}}{E^\varphi{}}\right)}{2 \beta  \sqrt{\Delta } E^x{}}-\frac{E^\varphi{} \sin
   ^2\left(\frac{\beta  \sqrt{\Delta } K_\varphi{}}{\sqrt{E^x{}}}\right)}{4 \beta ^2 \Delta
   \sqrt{E^x{}}}\nonumber\\
&+&\frac{E^\varphi{} K_\varphi{} \sin \left(\frac{\beta  \sqrt{\Delta } K_\varphi
   }{\sqrt{E^x{}}}\right) \cos \left(\frac{\beta  \sqrt{\Delta } K_\varphi{}}{\sqrt{E^x{}}}\right)}{2
   \beta  \sqrt{\Delta } E^x{}},\label{pde1}\\
\partial_t K_\varphi&=&\frac{\left(\partial_x E^x\right)^2}{8
   \sqrt{E^x{}} E^\varphi{}^2}-\frac{\sqrt{E^x{}} \sin \left(\frac{\beta  \sqrt{\Delta } K_\varphi{}}{\sqrt{E^x{}}}\right) \sin \left(\frac{2 \beta  \sqrt{\Delta } \sqrt{E^x{}} K_x{}}{E^\varphi{}}\right)}{\beta ^2 \Delta }\nonumber\\
   &+&\frac{2 E^x{} K_x{} \sin \left(\frac{\beta  \sqrt{\Delta } K_\varphi{}}{\sqrt{E^x{}}}\right) \cos \left(\frac{2 \beta  \sqrt{\Delta } \sqrt{E^x{}} K_x{}}{E^\varphi{}}\right)}{\beta  \sqrt{\Delta } E^\varphi{}}-\frac{\sqrt{E^x{}} \sin ^2\left(\frac{\beta  \sqrt{\Delta
   } K_\varphi{}}{\sqrt{E^x{}}}\right)}{2 \beta ^2 \Delta }-\frac{1}{2
   \sqrt{E^x{}}},\label{pde2}\\
\partial_t E^x&=&\frac{2 E^x{} \sin \left(\frac{\beta  \sqrt{\Delta } K_\varphi{}}{\sqrt{E^x{}}}\right) \cos \left(\frac{2 \beta  \sqrt{\Delta } \sqrt{E^x{}} K_x{}}{E^\varphi{}}\right)}{\beta  \sqrt{\Delta }},\label{pde3}\\
\partial_t E^\varphi&=&\frac{E^\varphi{} \cos \left(\frac{\beta 
   \sqrt{\Delta } K_\varphi{}}{\sqrt{E^x{}}}\right) \sin \left(\frac{2 \beta  \sqrt{\Delta }
   \sqrt{E^x{}} K_x{}}{E^\varphi{}}\right)}{\beta  \sqrt{\Delta }}+\frac{E^\varphi{} \sin
   \left(\frac{\beta  \sqrt{\Delta } K_\varphi{}}{\sqrt{E^x{}}}\right) \cos \left(\frac{\beta  \sqrt{\Delta
   } K_\varphi{}}{\sqrt{E^x{}}}\right)}{\beta  \sqrt{\Delta }}\label{pde4}.
\ee
when $E^x,E^\varphi>0$ are assumed.

The polymer quantization of the phase space $\calp$ is carried out in \cite{Zhang:2021xoa}, where the Hamiltonian ${\bf H}_\Delta$ is quantized on an 1-dimensional lattice along the $x$-direction. The U(1) holonomy $e^{i\l \int\rmd x A_1}$ and the point holonomy $e^{i\mu A_2}$ are among the basic variables in the quantization. The matrix elements of the time-evolution operator $e^{-\frac{iT}{\hbar}\hat{\bf H}_\Delta}$ is expressed as a phase space path integral:
\be
\langle\psi_1|e^{-iT\hat{\bf H}_\Delta}|\psi_2\rangle=\int D\mu[K_x,K_\varphi,E^x,E^\varphi]\, e^{\frac{i}{\hbar}S_\Delta[K_x,K_\varphi,E^x,E^\varphi]}.
\ee
The detailed expressions of the action $S_\Delta[K_x,K_\varphi,E^x,E^\varphi]$ and the path integral measure $D\mu$ can be found in \cite{Zhang:2021xoa}. The effective dynamics \eqref{pde1} - \eqref{pde4} is reproduced at the leading order in the $\hbar$-expansion of the path integral formula by the stationary phase approximation. Here $\Delta$ is viewed as independent from $\hbar$ in the expansion, although $\Delta\sim G\hbar\b$. This indicates that from the path integral point of view, the validity regime of the effective dynamics is given by scaling $\hbar$ small and $\b$ large while keeping $\Delta$ fixed. This regime is similar to the earlier path integral approach in LQC \cite{Ashtekar:2010gz}.

\section{Homogeneous reduction of the effective dynamics and its phenomenological implications}
\label{sec:homogeneous-collapse}
\renewcommand{\theequation}{3.\arabic{equation}}\setcounter{equation}{0}

In this section, we discuss the homogeneous reduction of our loop quantized model for the collapsing LTB spacetime and conclude that our model can reduce to the well-known  Oppenheimer-Snyder (OS) model with effective dynamics coinciding with $\bar{\mu}$-LQC for the homogeneous collapsing dust cloud with a non-vanishing dust energy density \cite{Giesel:2021dug}. For both the marginally bound case and the bound case, we analyze in detail the reduction ansatz, the dynamical equations as well as the formation of the trapped surfaces at the boundary in the resulting OS model.

\subsection{LTB dust shell model  }
\label{sec:RevLTBdustshell}

One approach to quantize LTB models is to start with a spherically symmetric spacetime, gauge-fix the Gauss constraint at the classical level and then apply LTB conditions that specialize the classical spherically symmetric spacetime to an LTB form and use this as the classical model for the quantization. This strategy was for instance followed in \cite{Kiefer:2019csi} and  \cite{Giesel:2021dug}. The work in \cite{Kiefer:2019csi} presents a Schrödinger quantization of model describing the dynamics of outermost dust shell for a homogeneous dust energy density and \cite{Giesel:2021dug} considers such a model for a different kind of loop quantization at the effective level. Both works restrict their discussion to the marginally bound case. In this subsection we want to  briefly review the main properties of the model \cite{Giesel:2021dug} because in the later part of this work we want to compare it with the results obtained from the path integral formalism presented here. We assume that the Gauss constraint has already been gauge-fixed and consider the metric in \eqref{metric1} as well as the Poisson bracket in \eqref{Poisson1} as the starting point. The LTB conditions for the marginally bound case in Ashtekar-Barbero variables read\footnote{Note that compared to \cite{Bojowald:2008ja} and \cite{Giesel:2021dug} the Poisson bracket in \eqref{Poisson1} involves an additional factor of $\frac{1}{2}$ and this results in an additional factor of $2$ in the second LTB condition in the notation used in this work.}
\begin{equation}
\label{mb LTB conditions}
{\rm I.}\,\,\, 2E^{\varphi}(t,x) - |\partial_x E^x|(t,x)\approx 0\quad\quad{\rm II.}\,\,\,
\partial_x K_{\varphi} {} (t,x) - 2 {\rm sgn}(E^x)K^x(t,x)\approx 0.
\end{equation}
As can be seen from \eqref{eq:ClassDiffeo} the combination of the LTB conditions in I and II together with the condition $P_x=0$ in \eqref{eq:ClassDiffeo} yield a vanishing contribution to the spatial diffeomorphism constraint at the classical level. As a consequence, the Brown-Kuchar dust model considered in \cite{Kiefer:2019csi,Giesel:2021dug} reduces to the case of non-rotational dust where as in the case of the Gaussian dust model the lapse is given by $N=1$ and the shift vector vanishes. For this reason it is reasonable to compare the model in \cite{Giesel:2021dug} with the results obtained here using Gaussian dust. Because the LTB conditions are applied in the classical model in \cite{Giesel:2021dug} one is left with one set of canonical variables only, that is $(K_x(t,x),E^x(t,x))$. As shown in \cite{Kiefer:2019csi}, if one imposes the assumption that the individual shells decouple at the classical level, an action for the outermost shell can be derived. In \cite{Giesel:2021dug} this shell model was used in connection variables and following their notation we denote the densitized triad of the outermost shell by $\widetilde{E}^x(t):=E^x(t,x_s)$, where $x_s$ in the radial coordinate of the shell and the conjugate connection variable by $\widetilde{K}_x(t):=K_x(t,x_s)$ that satisfy $\{\widetilde{K}_x,\widetilde{E}^x\}=G$ and whose classical dynamics is described by the following action
\begin{equation}
S=\frac{1}{G} \int \mathrm{d} \tau L_{\text {shell }}:=-\frac{1}{8 G} \int \mathrm{d} \tau \frac{(\frac{d\widetilde{E}^{x}}{d\tau})^{2}}{\sqrt{|\widetilde{E}^{x}|}}\quad {\rm with} \quad
H_s=-\frac{2}{G} \widetilde{K}_{x}^{2} \sqrt{|\widetilde{E}^{x}|}=-m_s,
\end{equation}
where $H$ denotes the corresponding physical Hamiltonian and $m_s$ stands for the dust mass enclosed by the outermost dust shell. The work in \cite{Giesel:2021dug} considers the usual loop quantization of the shell model based on holonomies and triads as well as a loop quantization involving in addition gauge covariant fluxes \cite{Liegener:2019ymd}. As we will not consider gauge covariant fluxes in our work here in the further discussion we will briefly summarise the effective model where gauge covariant fluxes are absent. The effective Hamiltonian involving holonomy corrections reads \cite{Giesel:2021dug}
\begin{equation}
H_{s}^{\Delta}=-\frac{(\widetilde{E}^{x})^{3 / 2}}{2 G \Delta \beta^{2}} \sin ^{2}\left(\frac{2 \beta\sqrt{\Delta} \widetilde{K}_{x}}{\sqrt{\widetilde{E}^{x}}}\right)=-m_s,
\end{equation}
with $\Delta=4 \sqrt{3} \pi \beta l_{p l}^{2}$ fixed by the minimum non-zero eigenvalue of area operator in LQG.
The corresponding equations of motion take the form
\begin{equation}
\begin{aligned}
\dot{\widetilde{E}^{x}} &=\frac{\widetilde{E}^{x}}{\beta \sqrt{\Delta}} \sin \left(\frac{4 \sqrt{\Delta} \beta \widetilde{K}_{x}}{\sqrt{\widetilde{E}^{x}}}\right), \\
\dot{\widetilde{K}_{x}} &=\frac{\widetilde{K}_{x}}{2 \sqrt{\Delta} \beta} \sin \left(\frac{4 \sqrt{\Delta} \beta k_{x}}{\sqrt{\varepsilon^{x}}}\right)-\frac{3 \sqrt{\widetilde{E}^{x}}}{4 \beta^{2} \Delta} \sin ^{2}\left(\frac{2 \sqrt{\Delta} \beta \widetilde{K}_{x}}{\sqrt{\widetilde{E}^{x}}}\right).
\end{aligned}
\end{equation}
Taking into account that $\widetilde{R}=\sqrt{|\widetilde{E}^x|}$ one can derive an effective equation for $\dot{\widetilde{R}}/\widetilde{R}$ yielding a modified Friedmann equation given by
\begin{equation}
\label{eq:ModFriedmann}
\left(\frac{\dot{\widetilde{R}}}{\widetilde{R}}\right)^{2}=\frac{8 \pi G}{3} \rho\left(1-\frac{\rho}{\rho_{\max }^{0}}\right)
\end{equation}
with $\rho=\frac{3 m_s}{4 \pi \tilde R^{3}}$ and where $\rho^{0}_{\mathrm{max}}=3 /\left(8 \pi G \beta^2 \Delta\right)$ denotes the maximum energy density enclosed by the outermost dust shell that is allowed in this model. That \eqref{eq:ModFriedmann} corresponds to a quantum gravity modified Friedmann equation with zero spatial curvature ($k=0$) reflects again the fact that the model corresponds to the marginally bound case.
The effective dynamics for the homogeneous dust collapse involves a quantum geometric correction term causing the right hand side of this equation to vanish when the density of the dust cloud reaches its maximum $\rho^{0 }_{\mathrm{max}}$. The numerical results in \cite{Giesel:2021dug} show that for a generic set of chosen initial conditions the singularity at $\widetilde{R}=0$ is replaced by a symmetric bounce. Furthermore, in the $k=0$ case, independent of the initial conditions, there exists a threshold for the dust mass below which no trapped surfaces will form in the dust collapse. In case the dust mass is larger than this threshold mass, then a pair of a dynamical black and white hole forms symmetrically around the bounce. Because in the model in \cite{Giesel:2021dug} the LTB conditions are implemented at the classical level where these are stable under the classical evolution \cite{Bojowald:2008ja} and one only considers the LTB canonical pair $(K_x,E^x)$, the LTB sector is preserved by construction. In contrast to above analysis, in this work here we consider the effective spherically symmetric model obtained from the path integral formalism in \cite{Han:2020uhb} and therefore going to the LTB sector requires corresponding LTB conditions to be implemented. In general these are not stable under the effective dynamics and for this reason the strategy followed in \cite{Bojowald:2008ja} is to modify the LTB conditions by additional functions depending on the triads chosen such that the stability is  ensured and the geometric part of the spatial diffeomorphism constraint is 
vanishing. As shown below, in the homogeneous reduction the corresponding LTB conditions are stable under the effective dynamics of the model in this work. More general  will be considered elsewhere \cite{GLRSWToAppear}. 

\subsection{A homogeneous reduction to the Oppenheimer-Snyder dust collapsing model}
\label{sec:OSdustmodel}

The Oppenheimer-Snyder (OS) model describes the gravitational collapse of a homogeneous matter cloud  whose interior spacetime is isometric to the cosmological spacetimes and correspondingly its metric is  given by
\begin{eqnarray}
\label{os metric}
d s^{2}=-d t^{2}+\frac{a(t)^{2}}{1-k x^{2}} d x^{2}+x^{2} a(t)^{2} d \Omega^{2},
\end{eqnarray} 
where $a(t)$ denotes the scale factor and the constant $k$ is used to describe two distinct cases with $k=0$ corresponding to the marginally bound case and $k=1$ to the bound case. 
Comparing this metric with \eqref{metric1}, we obtain
\begin{eqnarray}
\label{homo_metric_an}
E^x(t,x) = x^2 a(t)^2, \qquad E^\varphi(t,x) = \frac{x a(t)^2}{\sqrt{1- k x^2}} \,,
\end{eqnarray}
which satisfy the LTB condition \cite{Bojowald:2008ja,Giesel:2021dug}
\begin{eqnarray}
\label{LTB1}
E^x {} '(t,x) = 2 \sqrt{1- k x^2} E^\varphi(t,x).
\end{eqnarray}
Here without loss of generality we assume $E^{x} \geq 0$ and (\ref{LTB1}) reduces to (\ref{mb LTB conditions}) for the marginally bound case. The results for $E^{x} < 0$ can be obtained from the symmetry of the equations of motion \eqref{pde1}-\eqref{pde4}. Furthermore, Eq. (\ref{homo_metric_an}) can be regarded as a homogeneous reduction of the variables  $E^x(t,x) $ and $E^\varphi(t,x)$ as their spatial dependence is clearly spelled out.  Therefore, the only dynamical variable is the scale factor which is a constant at any comoving radius $x$  at a given time.

A corresponding homogeneous reduction of the conjugate momenta $K^x(t,x) $ and $K^\varphi(t,x)$ can be derived by requiring the consistency of the equations of motion of $E^x(t,x) $ and $E^\varphi(t,x)$. Plugging \eqref{homo_metric_an} into Eqs. \eqref{pde3}-\eqref{pde4}, these two equations reduce to 
\begin{eqnarray}
\frac{\dot a(t)}{a(t)} &=&\frac{ \sin \left(\frac{\beta  \sqrt{\Delta } K_{\varphi}(t,x)}{x a(t)}\right) \cos \left(\frac{2 \beta  \sqrt{\Delta(1- k x^2)} K_x(t,x)}{a(t)}\right)}{\beta  \sqrt{\Delta }} \,, \label{eom_eq3}\\
\frac{\dot a(t)}{a(t)}&=&\frac{\cos \left(\frac{\beta  \sqrt{\Delta } K_{\varphi}(t,x)}{x a(t)}\right) \left(\sin \left(\frac{2 \beta  \sqrt{\Delta( 1 -  k x^2)} K_x(t,x)}{a(t)}\right)+\sin \left(\frac{\beta  \sqrt{\Delta } K_{\varphi}(t,x)}{x a(t)}\right)\right)}{2 \beta  \sqrt{\Delta }},\label{eom_eq4}
\end{eqnarray} 
where $\dot{a}(t):=\frac{da(t)}{dt}$.
Since $a(t)$ only depends on the time coordinate, one can impose the following ansatz:
\begin{eqnarray}
\label{ansatz2}
K_{\varphi}(t,x) = x \tilde{K}_{\varphi}(t), \qquad K_{x}(t,x) = \frac{\tilde{K}_{x}(t)}{\sqrt{1- k x^2}},
\end{eqnarray} 
so that the arguments of the trigonometric functions only depend on $t$. The consistency between \eqref{eom_eq3} and \eqref{eom_eq4} requires
\begin{eqnarray}
-\sin \left(\frac{\beta  \sqrt{\Delta } (\tilde{K}_{\varphi}(t)+2 \tilde{K}^{x}(t))}{a(t)}\right)-3 \sin \left(\frac{\beta  \sqrt{\Delta } (\tilde{K}_{\varphi}(t)-2 \tilde{K}^{x}(t))}{a(t)}\right)+\sin \left(\frac{2 \beta  \sqrt{\Delta } \tilde{K}_{\varphi}(t)}{a(t)}\right) =0.\nb\\
\end{eqnarray} 
One set of the solutions to this constraint is given by 
\begin{eqnarray}
\tilde{K}_{\varphi}(t) = 2 \tilde{K}^{x}(t) + \frac{2\pi n a(t)}{\beta \sqrt{\Delta}} \,, \qquad n \in \mathbb{Z}.
\end{eqnarray}
In order to fix the parameter $n$, we plug the ansatz (\ref{ansatz2}) into the  equations of motion of  $K^x(t,x) $ and $K^\varphi(t,x)$, namely \eqref{pde1} and \eqref{pde2}, and find the consistency condition of the resulting equations demands $n = 0$. As a result, in the ansatz (\ref{ansatz2}), we also require
\begin{eqnarray}
\label{ansatz3}
\tilde{K}_{\varphi}(t) = 2 \tilde{K}^{x}(t) \,,
\end{eqnarray}
which implies the fulfillment of the classical LTB condition  \cite{Bojowald:2008ja,Giesel:2021dug}, i.e.
\begin{eqnarray}\label{LTB2}
 \partial_x K_{\varphi} {} (t,x) = 2 \sqrt{1- k x^2} K^x(t,x),
\end{eqnarray}
at the level of the effective dynamics for the collapse of a homogeneous dust cloud.
Based on the above analysis, we conclude that in the homogeneous reduction with the ansatz (\ref{homo_metric_an}), (\ref{ansatz2}) and (\ref{ansatz3}), the collapse of an inhomogeneous dust cloud whose dynamics is governed by \eqref{pde1}-\eqref{pde4} can be reduced to the collapse of a homogeneous dust cloud with the following reduced equations of motion
\begin{eqnarray}
\label{reduced equations of motion}
\dot a(t) = \frac{a(t) \sin \left(2 \beta  \sqrt{\Delta } b(t)\right)}{2 \beta  \sqrt{\Delta }},\quad \dot b(t) =- \frac{1}{2} \left(\frac{k}{a(t)^2}+\frac{3 \sin ^2\left(\beta  \sqrt{\Delta } b(t)\right)}{\beta ^2 \Delta }\right),
\end{eqnarray}
where we have defined 
\begin{equation}
b(t) := \frac{2 \tilde{K}_{x}(t)}{a(t)}, ~~ v:=a^3(t). 
\end{equation}
In the next subsection, one can find that the above equations of motion could be derived from the effective Hamiltonian density resulting from the homogeneous reduction. 

\noindent 
{\bf Remark 1}: Note that in general, the action of the Hamiltonian on the LTB condition \eqref{LTB1} gives
\begin{eqnarray}\label{H_LTB}
&&\Big\{{\bf H}_{\Delta} \, , \, E^{\varphi}(x) - \frac{\partial_x E^x(x)}{2\sqrt{1-k x^2}} \Big\} = - \frac{1}{\beta   \left(\Delta +\Delta  k x^2\right)^{3/2}} \Bigg\{ \Delta 
E^{\varphi}(x) \left(1 - k x^2\right)^{3/2} \nonumber\\
&& \cos \left(\frac{\beta  \Delta  K_{\varphi}(x)}{\sqrt{\Delta  E^x(x)}}\right) 
 \left(\sin \left(\frac{2 \beta  K_x(x) \sqrt{\Delta  E^x(x)}}{E^{\varphi}(x)}\right)+\sin \left(\frac{\beta  \Delta  K_{\varphi}(x)}{\sqrt{\Delta  E^x(x)}}\right)\right) \\
&& + \Delta ( 3 k x - (1-k x^2) \partial_x ) \Big[\cos \left(\frac{2 \beta  K_x(x) \sqrt{\Delta  E^x(x)}}{E^{\varphi}(x)}\right)E^x(x)\sin \left(\frac{\beta  \Delta  K_{\varphi}(x)}{\sqrt{\Delta  E^x(x)}}\right) \Big]
\Bigg\} . \nonumber
\end{eqnarray}
This implies that the LTB condition \eqref{LTB1} is generally not preserved by the effective dynamics after the system is polymerized with the $\bar \mu$ scheme. However, in the particular case of the homogeneous reduction, with the ansatz \eqref{homo_metric_an}, \eqref{ansatz2} and (\ref{ansatz3}), the right-hand side of the above equation identically vanishes, thus LTB condition \eqref{LTB1} is preserved. Similar analysis can be carried out with respect to the LTB condition (\ref{LTB2}) which is found to be preserved as well during the evolution of the homogeneous dust cloud. Whereas for the evolution of the inhomogeneous dust cloud, since the right-hand side of the above Poisson bracket  does not vanish,  the classical LTB conditions are no longer preserved. As a result, one is required to find the analogues of the classical LTB conditions for the polymerized system which was addressed in \cite{Bojowald:2008ja}, but this possible generalization is 
beyond the scope of the current study. A more detailed investigation on the LTB conditions will be considered in future work \cite{GLRSWToAppear}.

\subsection{The Hamiltonian and the evolutionary properties of the interior of the homogeneous dust collapse}
\label{sec:hamiltonian-and-conserved-quantity}

After the homogeneous reduction with the ansatz \eqref{homo_metric_an}, \eqref{ansatz2} and (\ref{ansatz3}), it can be shown in a straightforward way that the Hamiltonian density $\cc_\Delta(x)$ in \eqref{Hamiltonian density} reduces to 
\begin{eqnarray}
\label{effective Hamiltonian from reduction}
\cc_\Delta(x) &=& - \frac{24 \pi a(t)^3  x^2 }{\kappa \sqrt{1- k x^2} }\left(  \frac{ \sin ^2\left(\beta  \sqrt{\Delta } b(t)\right)}{\beta ^2 \Delta }  + \frac{  k }{a(t)^2 } \right)\nb\\
&=& - \frac{12 \pi a(t)^3 }{\kappa } \partial_x \left( \frac{x \sqrt{1 - k x^2}}{ k}-\frac{\sinh ^{-1}\left(\sqrt{k} x\right)}{ k^{3/2}} \right)\left( \frac{ \sin ^2\left(\beta  \sqrt{\Delta } b(t)\right)}{\beta ^2 \Delta    }  +\frac{  k }{a(t)^2} \right) \,.
\end{eqnarray}
The corresponding classical limit of the Hamiltonian density can be recovered as $\Delta \to 0$, leading to
\begin{eqnarray}\label{classical_ham_homo}
\cc_\mathrm{classical} = \lim_{\Delta \to 0}\cc_\Delta(x) =- \frac{24 \pi  x^2 a(t) \left(b(t)^2 a(t)^2 + k\right)}{\kappa  \sqrt{1-k x^2}} =- \frac{24 \pi  x^2 a(t) \left(\dot a(t)^2 +k\right)}{\kappa  \sqrt{1-k x^2}},
\end{eqnarray}
here we have used $b=\dot a/a$ obtained from the equation of motion of $\dot a$ in (\ref{reduced equations of motion}) in the classical limit.
$\cc_\mathrm{classical}$ is exactly the classical Hamiltonian for the gravitational collapse of a homogeneous dust cloud in the marginally bound case ($k=0$)  \cite{Bojowald:2008ja} or the bound case ($k=1$) \cite{Bojowald:2009ih}.
On the other hand, with the dust energy density given by
\begin{eqnarray}
\label{energy density}
\rho = - \frac{\cc_{\Delta}}{4 \pi E^{\varphi} \sqrt{E^x}} =\frac{6}{{\kappa } } \left(\frac{\sin ^2\left(\beta  \sqrt{\Delta } b(t)\right)}{\beta ^2 \Delta }+\frac{k}{a(t)^2}\right) \, ,
\end{eqnarray}
assuming the comoving radius of the outermost dust shell is denoted by $x_s$,  the dust mass $m_s$ enclosed within the dust cloud turns out to be 
\begin{eqnarray}
\label{dust mass}
m_s = \frac{4 \pi x_s^3 a(t)^3 \rho}{3} .
\end{eqnarray}
Clearly, quantities proportional to $\epsilon := \rho a(t)^3$, e.g. the dust mass $m_s$, are
conserved.
Using the energy density defined in (\ref{energy density}) and the equation of motion of $a(t)$ in (\ref{reduced equations of motion}), one can obtain the following effective Friedmann equation
\begin{eqnarray}
\label{Friedmann from homogeneous reduction}
H^2 &=& \frac{\dot a(t)^2}{a(t)^2} = \left(\frac{\kappa  \rho }{6}-\frac{k}{a(t)^2} \right) \left( 1 -\frac{\beta ^2 \Delta  \kappa  \rho }{6}+\frac{\beta ^2 \Delta  k}{a(t)^2} \right) \nb\\
&=& \left(\frac{8 \pi G \rho }{3}-\frac{k}{a(t)^2} \right) \left( 1 -\frac{\beta ^2 \Delta  8 \pi G   }{3}\Big( \rho-\frac{3 k}{ 8 \pi G  a(t)^2} \Big) \right) \,,
\end{eqnarray}
which coincides with the modified Friedmann equation for the $K$ quantization of the spatially flat ($k=0$) or closed ($k=1$) FLRW universe with the $\bar \mu$ scheme \cite{Ashtekar:2006wn,2006PhDT1V}. As a result, a bounce with $\dot a(t_b)=0$ and $\ddot a(t_b)>0$ will take place at the time $t_b$ when $b(t_b) = \frac{\pi}{2 \beta \sqrt{\Delta}}$,
which once plugged into (\ref{energy density}) leads to
\begin{eqnarray}
\frac{6}{{\kappa } } \left(\frac{1}{\beta ^2 \Delta }+\frac{k}{a(t_b)^2}\right) a(t_b)^3 = \epsilon.
\end{eqnarray}
Therefore, the scale factor at the bounce can be solved from the above equation, yielding a general solution for any $k$
\begin{eqnarray}
\label{scale factor a}
a_b=a(t_b)=\frac{\left(3 \beta ^2 \Delta  \kappa  \epsilon +\sqrt{9 \beta ^4 \Delta ^2 \kappa ^2 \epsilon ^2+48 \beta ^6 \Delta ^3 k^3}\right)^{2/3}-2 \sqrt[3]{6} \beta ^2 \Delta  k}{6^{2/3} \sqrt[3]{3 \beta ^2 \Delta  \kappa  \epsilon +\sqrt{9 \beta ^4 \Delta ^2 \kappa ^2 \epsilon ^2+48 \beta ^6 \Delta ^3 k^3}}}.
\end{eqnarray}
Since the evolution of the dust cloud in the marginally bound case with $k=0$ is qualitatively different from that in the bound case with $k=1$. In the following, we  discuss these two cases separately. 

\noindent 
{\bf Remark 2}: It is important to note here two assumptions when dealing with the bound case. Being spatially-compact not only do quantum geometric effects enter via holonomies but also via inverse scale factor effects. Since the latter are generally negligible in the dynamics of the homogeneous and isotropic bounce they have been ignored \cite{Ashtekar:2006es}. Nevertheless their contribution can be significant in singularity resolution, such as in anisotropic cases \cite{Gupt:2011jh}. The second assumption is that the $K$ quantization is based on constructing holonomies using the extrinsic curvature only. Since we are working the approximation where inverse scale factor effects are ignored and intrinsic curvature does not enter the holonimies, above modified Friedmann equation \eqref{Friedmann from homogeneous reduction} for the bound case ignores quantum geometric modificiations to the intrinsic curvature. 
The following analysis for the bound case would be under these setting and it is an open question how the results change if these assumptions are relaxed. \\

\vskip0.5cm

{\bf Case A: The marginally bound case} 

In this case $k=0$, hence the scale factor at the bounce reads $a_b =\frac{\sqrt[3]{4 \beta ^2 \Delta  \kappa  \epsilon }}{2 \sqrt[3]{3 }}$, which once plugged into (\ref{energy density}) yields the maximum energy  density  at the bounce, namely 
\begin{eqnarray}
\label{maximum energy density}
\rho^{0}_\mathrm{max}= \frac{6}{\kappa \beta ^2 \Delta  } =  \frac{3}{8 \pi G \beta ^2 \Delta  } \,.
\end{eqnarray}
For a dust cloud with a fixed mass $m_s$, it collapses continuously with a decreasing radius and an increasing energy density. When the energy density attains its maximum value at  $\rho^{0}_\mathrm{max}$, the bounce takes place and the dust cloud starts to re-expand towards spatial infinity. In this process, during the collapse of the dust cloud, $b$ lies in the interval $b\in( \frac{\pi}{2 \beta \sqrt{\Delta}}, \frac{\pi}{ \beta \sqrt{\Delta}})$ and  is continuously decreasing. After the quantum bounce when the dust cloud enters into expanding phase, $b$ monotonically decreases from $\frac{\pi}{2 \beta \sqrt{\Delta}}$ towards zero. In the marginally bound case, $b$ can not reach zero in any finite coordinate time. \\

{\bf Case B: The  bound case} 

In this case $k=1$, the quantum corrections enter into the effective Friedmann equation (\ref{Friedmann from homogeneous reduction}) in the second parenthesis on the right-hand side. Due to the spatial curvature, there is also a recollapse which takes place when the energy density satisfies $\rho_\mathrm{re}=\frac{3}{8\pi G a^2_\mathrm{re}}$, hereafter the index `re' will be used to denote quantities at the recollapse point. Combining with the relation between the energy density and the scale factor given in (\ref{dust mass}), it is straightforward to obtain the energy density and the scale factor at the recollapse which turn out to be 
\bq
a_\mathrm{re}=\frac{2Gm_s}{x^3_s},\quad \quad \rho_\mathrm{re}=\frac{3x^6_s}{32\pi G^3 m^2_s}.
\eq
On the other hand, in this case, the maximum energy density at the bounce takes the form
\bq
\label{med bound}
\rho_b=\rho^{0}_\mathrm{max}+\frac{3}{8\pi G a^2_b},
\eq
with $a_b$ given by (\ref{scale factor a}) for $k=1$. Therefore, for the bound case, the dust cloud behaves like a pulsating star which experiences infinite cycles of the bounces and the  recollapses with the energy density $\rho=\rho_b$ and  $\rho=\rho_\mathrm{re}$ respectively.

\subsection{The null expansions and the formation of the trapped surfaces of the homogeneous collapsing dust cloud}
\label{sec:null-expansion-trapped-region-inner-horizon}

In order to investigate the  formation of the trapped surfaces during the gravitational collapse of the dust cloud, for a generic spherically symmetric spacetime described by the metric \eqref{metric1}, one can define two future-directed null vectors which are normal to the sphere with the constant radius $\sqrt{|E^x|}=const$ via
\bq\label{null_expansion}
\partial_{\xi^+}=\frac{1}{\sqrt 2}\left(\partial_t+\frac{\sqrt{|E^x|}}{E^\varphi}\partial_x\right),\quad \quad \partial_{\xi^-}=\frac{1}{\sqrt 2}\left(\partial_t-\frac{\sqrt{|E^x|}}{E^\varphi}\partial_x\right).
\eq
If the radius of the sphere shrinks along the radial null geodesics $\xi^+=const$  and $\xi^-=const$, then a trapped surface forms at the sphere \cite{Hayward:1994bu}. In practice, it is convenient to introduce the expansion parameters $\theta_\pm$ which are defined by 
\bq
\theta_\pm=\frac{2}{\sqrt{|E^x|}}\partial_\pm\sqrt{|E^x|}=\frac{1}{\sqrt 2 |E^x|}\left(\partial_t|E^x|\pm\sqrt{|E^x|}\frac{\partial_x |E^x|}{E^\varphi}\right),
\eq
where $\partial_{\pm}$ denotes derivatives with respect to $\xi^{\pm}$ respectively. When $\theta_\pm<0$, the light rays emitted from the sphere converge on both sides of the sphere, then the sphere becomes a future trapped surface. In the homogeneous reduction with the ansatz (\ref{homo_metric_an}), the expansion parameters are simplified to 
\begin{eqnarray}
\theta_\pm=\frac{1}{\sqrt 2 xa}\left(x\dot a\pm\sqrt{1-kx^2}\right)=\frac{1}{\sqrt 2 R}\left(\dot R\pm\sqrt{1-kx^2}\right),
\end{eqnarray} 
where in  the last step we have used the definition of  the radius of the sphere $R\coloneqq xa(t)$ in the homogeneous case. 
Consequently, during the collapse of the dust cloud with $\dot{R}< 0$, a marginally trapped surface with $\theta_-<0$ and $\theta_+=0$ exists at the comoving radius 
\begin{eqnarray}
x_{h} = \frac{1}{\sqrt{\dot a^2+k}}.
\end{eqnarray} 
Note $R \, \theta_+$ decreases monotonically as the comoving radius $x$ increases. When the comoving radius of the outermost dust shell is larger than $x_h$, namely $x_s> x_h$, the outermost dust shell becomes a trapped surface with $\theta_\pm<0$. As a result, the criterion for the formation of the (marginally) trapped surface at the boundary of the dust cloud is 
\bq
R_s\ge R_h,
\eq
where $R_s= ax_s$ is the physical radius of the outermost dust shell and $R_h= ax_h$ denotes the physical radius of the marginally trapped surface located at $x=x_h$. 

The classical description of the collapse of the dust cloud can be obtained by taking the classical limit of the effective dynamics,  under which the dust mass in (\ref{dust mass}) tends to its classical value given by 
\bq
m^{\mathrm{c}}_s=\evalat[\Big]{m_s}{\Delta\rightarrow 0}=\frac{x^3_sa}{2G}\left(\dot a^2+k\right),
\eq
hereafter we use superscript $``c"$ to denote the quantities obtained in the classical limit when the minimal area gap tends to vanish.
As a result, in the classical theory, for both the marginally bound and the bound case, the physical radii of the marginally trapped surface and the outermost dust shell are related via
\begin{eqnarray}
R_{h}^{c}=x_{h}^{c} a(t) = \frac{x_s a(t)  \sqrt{x_s a(t)}}{\sqrt{2 G m^c_s}} = R_s  \sqrt{\frac{{R_s}}{{2G m^\mathrm{c}_s}}}.
\end{eqnarray}
Now imagine a dust cloud starts to collapse at a very large volume with $R_s\gg 2G m^\mathrm{c}_s$, at the early stage of the collapse, $R_{h}^{c}\gg R_s$, so its outermost shell is not trapped at all. As the dust cloud keeps collapsing, the critical moment  happens at  $R_s=2 G m^\mathrm{c}_s$ when the outermost dust shell becomes marginally trapped since $R_{h}^{c}=R_s$ at this moment. Afterwards, the outermost dust shell remains a  trapped  surface until the classical singularity at $R_s=0$ is reached. Therefore, in the classical case, the singularity is always covered by a trapped surface at the boundary of the dust cloud, namely an apparent horizon, which is consistent with the Cosmic Censorship Hypothesis. 

In contrast, assuming the validity of effective dynamics in the entire evolution (\ref{effective Hamiltonian from reduction}), a trapped surface at the boundary may not always form during the collapse of the dust cloud if the quantum bounce takes place before the formation of the trapped surface. In particular, as discussed above, there will be no trapped surface if  $R_{h} > R_{s}$, namely,
\begin{eqnarray}
\frac{R_{h}}{R_{s}}  =  \frac{1}{x_s \sqrt{\dot{a}^2 + k}} > 1 \, ,
\end{eqnarray}
holds for all the time before and after the bounce. Combining Eqs. (\ref{energy density})-(\ref{dust mass}) and the equations of motion in (\ref{reduced equations of motion}), it is straightforward to show that 
\bq
\label{critical ratio}
\frac{R_{h}}{R_{s}}=\frac{a^{1/3}}{(2 Gm_s)^{1/3}}\left(\frac{a^2\sin^2(\beta \sqrt \Delta b)}{\beta^2 \Delta}+k\right)^{1/3}\left(\frac{a^2\sin^2(2\beta \sqrt \Delta b)}{4\beta^2 \Delta}+k\right)^{-1/2}.
\eq
As a result, the minimum of $R_{h}/R_{s}$ is located at $\frac{d(R_{h}/R_{s})}{dt}=0$ which results in 
\begin{eqnarray}
4 a(t)^2 \sin ^3\left(\beta  \sqrt{\Delta } b(t)\right) \cos \left(3 \beta  \sqrt{\Delta } b(t)\right)+\beta ^2 \Delta  k \sin \left(4 \beta  \sqrt{\Delta } b(t)\right) = 0.
\end{eqnarray}
For the marginally bound case $k=0$, the above equation yields two solutions for $b$, namely, 
\begin{eqnarray}
\label{solution b}
b_1 = \frac{\pi }{6 \beta  \sqrt{\Delta }}\,, \quad b_2 =\frac{5 \pi }{6 \beta  \sqrt{\Delta }}.
\end{eqnarray} 
The first solution $b_1$ corresponds to the expanding phase of the dust cloud after the occurrence of the quantum bounce while the second solution $b_2$ to the collapsing phase of the dust cloud before the quantum bounce. These two solutions give the same minimum of $R_{h}/R_{s}$ as the evolution of the dust cloud is symmetric with respect to the bounce. Now plugging the solutions (\ref{solution b}) into the ratio (\ref{critical ratio}), one can immediately find the minimum of $R_{h}/R_{s}$ turns out to be 
\begin{eqnarray}
\evalat[\Bigg]{ \frac{R_{h}}{R_s}}{min}= \left(\frac{ 8 \sqrt{\Delta} \beta}{3\sqrt{3}\, G m_s}  \right)^{1/3} .
\end{eqnarray} 
Then we can find a threshold mass for the formation of the trapped surface at the boundary of the dust cloud, which is 
\begin{eqnarray}
M_*=\frac{ 8 \sqrt{\Delta}  \beta}{3\sqrt{3} G}.
\end{eqnarray}
When the dust mass $m_s$ is less than $M_*$, $R_{h}/R_{s}$ is always larger than unity during the entire evolution of the dust cloud which implies no trapped surface (horizon) would form at any time. Only when $m_s$ is taken to be larger than $M_*$, the horizon can form before the occurrence of the bounce during the collapse of the dust cloud. More details on the qualitative features of dynamical evolution of the dust cloud and the formation of the trapped surface will be discussed and analyzed via numerical simulations in the next section.

\noindent
{\bf Remark 3:} It is worthwhile to note that our results on the marginally bound case with $k=0$ are consistent with those reported in \cite{Giesel:2021dug}. In particular, the threshold mass $M_*$ is exactly the same as the one derived in \cite{Giesel:2021dug}. Moreover, the numerical results presented in Sec. \ref{sec:numerics} will further confirm this consistency. Although we expect that in more general models implementing the LTB conditions and quantization will not commute as in the case of the relationship between polymerization and the gauge fixing recently discussed in \cite{Giesel:2021rky}, for the polymerization at the level of the effective dynamics commutes with the homogeneous reduction at least for the $K$ quantization with the $\bar \mu$ scheme. In particular, we have shown this explicitly for the marginally bound case as  we have obtained from the homogeneous reduction the same modified dynamical equations and the threshold mass for the formation of the trapped surface at the boundary as in the dust shell model which relies on a loop quantization of the classical homogeneous model of the dust collapse \cite{Giesel:2021dug}. 

\subsection{The exterior stationary spacetime and the matching conditions}
\label{sec:MatchingCond}

To explore potential phenomenological signatures of collapse of the dust cloud, it is necessary to glue the interior collapsing spacetime with an exterior spacetime which describes the geometry and thus the matter distributions of the collapsing dust cloud as  observed by an outside spectator at spatial infinity. As discussed in Sec. \ref{sec:OSdustmodel}, the interior collapsing spacetime is described  by the OS model in the classical regime  while for the exterior spacetime  we choose without loss of generality a generic spherically symmetric spacetime with its metric given in (\ref{metric1}). The matching is performed at the boundary $x=x_s$. In particular, for the interior spacetime, its first and second fundamental forms on the boundary surface turn out to be
\begin{eqnarray}
\gamma_{\mu \nu}^{-} dx^{\mu}dx^{\nu} &=& -dt^2 + x_s^2 a(t)^2 d \Omega^2,\\
K_{\mu \nu}^{-}dx^{\mu}dx^{\nu} &=& \frac{1}{2} \partial_{x} E^{x} \sqrt{\frac{{E^{x}}}{(E^{\varphi}){}^2}} d \Omega^2 = x_s a(t) \sqrt{1 - k x_s^2}  d \Omega^2,
\end{eqnarray}
where $\gamma_{\mu\nu}^{-}$ is the induced 3-metric on the boundary surface and $K^-_{\mu \nu} = - \gamma_{\mu}^{\alpha}\gamma_{\nu}^{\beta} \nabla_{\alpha} n_{\beta}$ is the projection of the extrinsic curvature onto the boundary surface.
While for the exterior solution, assuming the boundary surface $\Sigma$ is determined by $F(\tau,x):= f(\tau)-x =0$, 
where the normal co-vector is given by $n_{\mu} = \nabla_{\mu} F(\tau,x)$ which satisfies $n_{\mu} n^{\mu} > 0$, 
then the induced metric and extrinsic curvature projected from the exterior spacetime onto the boundary surface take the form 
\begin{eqnarray}\label{junction_ltb}
\gamma_{\mu \nu}^{+}dx^{\mu}dx^{\nu}  &=& - dt^2 + E^x(\tau(t), x(t)) d \Omega^2,\\
K_{\mu \nu}^{+}dx^{\mu}dx^{\nu} &=& 
B d t^2 + C  d \Omega^2,
\end{eqnarray}
with
\begin{eqnarray}
B &=&\frac{1}{2 E^x \text{sgn}(E^{\varphi}) \left(E^x-f'(\tau)^2 (E^{\varphi})^2\right)^{3/2}} \left( f'(\tau) E^{\varphi} E^x \left(f'(\tau) \left(2 f'(\tau) E^{\varphi} \partial_\tau E^{\varphi}+\partial_x E^{x}\right)+2 \partial_\tau E^{x}\right) \right. \notag\\
&&\qquad \qquad \left. -2 (E^x)^2 \left(f'(\tau) \left(f'(\tau) \partial_x E^{\varphi}+2 \partial_\tau E^{\varphi}\right)+f''(\tau) E^{\varphi}\right)+f'(\tau)^3 (E^{\varphi})^3 \partial_\tau E^{x} \right), \,  \\
C &=&\frac{f'(\tau) (E^{\varphi})^2 \partial_\tau E^{x}+E^x \partial_x E^{x}}{2 \left| E^{\varphi}\right|  \sqrt{E^x-f'(\tau)^2 (E^{\varphi})^2}} . 
\end{eqnarray}
Requiring the continuation of the induced metric and the existence of a surface stress-energy tensor on the boundary surface $\Sigma$, the  matching conditions of the interior and the exterior spacetimes are prescribed by
\begin{eqnarray}\label{junction_condition}
\gamma_{\mu \nu}^{+}  -   \gamma_{\mu \nu}^{-} & = &0,\\
( K_{\mu \nu}^{+} - \gamma_{\mu\nu}^{+} K^{+}) -   (K_{\mu \nu}^{-} - \gamma_{\mu\nu}^{-} K^{-})& =& \sigma_{\mu \nu},
\end{eqnarray}
where $\sigma_{\mu \nu}$ stands for the surface stress-energy tensor on $\Sigma$. Note the exact form of $\sigma_{\mu \nu}$ is determined by the specific exterior spacetime metric used to match with the interior. Taking the classical Schwarzschild exterior as an example, we have $E^x= R^2, E^{\varphi} =R' R$ with 
\begin{eqnarray}\label{sch_sol}
R(x,\tau) = \left( \frac{3}{2} \sqrt{{2 m}} (x - \tau) \right)^{\frac{2}{3}} =\left( \frac{3}{2} \sqrt{2{ m}} z \right)^{\frac{2}{3}}.
\end{eqnarray}
With the classical homogeneous interior solutions  given by
\begin{eqnarray}\label{classical_homo_sol}
a(t) = \left(\frac{3}{2} \sqrt{{2 \mathcal{ E}}}(t_0-t) \right)^{\frac{2}{3}} \quad \mathrm{for} \quad k=0,
\end{eqnarray}
the junction condition (\ref{junction_condition}) in the classical marginally bound case can be solved with $\sigma_{ab} = 0$ and
\begin{eqnarray}\label{junction_classical}
E^x = x_s^2 a(\tau)^2, \; m = x_s^3 \mathcal{ E}, \; x(t) = x_s, \; \tau(t) = t - x_s \, . 
\end{eqnarray}

\subsubsection{The effective stationary exterior solution}
\label{sec:EffStatexSol}
To obtain an analog of the Schwarzschild solution \eqref{sch_sol} in the effective dynamics, we can introduce the following generator of the Killing vector field
\begin{eqnarray}\label{killing}
   \partial_{K} = \partial_t + \partial_x,
\end{eqnarray}
so that the metric functions $E^x, E^{\varphi}$ are preserved by $\partial_{K}$. As a result, we have the following ansatz
\be
&&E^x(t,x)=E^x(z),\quad E^\varphi(t,x)=E^\varphi(z),\nonumber\\
&& K_x(t,x)=K_x(z),\quad K_\varphi(t,x)=K_\varphi(z),\quad z=x-t.\label{ansztzz}
\ee
With the null expansion \eqref{null_expansion}, one can easily check that, the vector field $\partial_{K}$ is timelike in the untrapped region, while spacelike inside the trapped region. Thus $\partial_{K}$ generates the analog of classical static solution in the effective dynamics. This solution has been studied in detail in \cite{Han:2020uhb}. We will briefly summarize the results here.

With ansatz \eqref{ansztzz}, the EOMs \eqref{pde1} - \eqref{pde4} reduce to a set of 1st order ODEs:
\be
\frac{\rmd}{\rmd z}\left[\begin{array}{c}E^x \\E^\varphi  \\K_1  \\K_2 \end{array}\right]= - \left[\begin{array}{c}f^x\lt(E^x,E^\varphi,K_1,K_2\rt)\\ f^\varphi\lt(E^x,E^\varphi,K_1,K_2\rt)\\ f_1\lt(E^x,E^\varphi,K_1,K_2\rt)\\ f_2\lt(E^x,E^\varphi,K_1,K_2\rt)\end{array}\right].\label{ODE1}
\ee
The classical solution is supposed to be recovered at $z\to+\infty$ or $z \to -\infty$, since it is far away from the classical singularities $z=0$. The Schwarzschild solution \eqref{sch_sol} then can be given at $z \gg 1$ or $z \ll -1$ as initial conditions for the ODEs \eqref{ODE1}. 

Note that, using coordinate $z$, we can rewrite the metric for black hole exterior ($\partial_{K}$ timelike) as 
\bq
\label{metric1z}
\rmd s^2=-\rmd t^2+\frac{(E^\varphi(z))^2}{|E^x(z)|} (\rmd z+\rmd t)^2+|E^x|\rmd\Omega^2.
\eq
By defining a new coordinate $ \tau = t- \int_{z_0}^z dz' \frac{E^{\varphi}(z')^2 }{ E^{\varphi}(z')^2 - E^x(z')}$, the above metric turns out to be  equivalent to 
\begin{equation}\label{metrictz}
    \rmd s^2 = - \frac{E^x -(E^{\varphi})^2}{E^x}(z) \rmd \tau^2 + \frac{(E^{\varphi})^2 }{E^x -(E^{\varphi})^2}(z) \rmd z^2 + E^x(z) \rmd \Omega^2,
\end{equation}
which has coordinate singularities at the horizons located at $E^x =(E^{\varphi})^2$. Moreover, 
in the region where $E^x(z)$ is monotonic, using $E^x(z)=R^2$ it  can be further rewritten as
\begin{equation}
\label{metricz_new}
    \rmd s^2 = - \lt(1 + \frac{(E^{\varphi})^2}{R^2}\rt) \rmd \tau^2 + \frac{(E^{\varphi})^2 }{(R^2 -(E^{\varphi})^2) (R')^2} \rmd R^2 + R^2 \rmd \Omega^2.
\end{equation}
Therefore, the above coordinate transformation has a coordinate singularity at the bounce point $R'=0$.

\textbf{Remark 4:} Note that the above static solution may not  be a vacuum solution for given vacuum initial values on the initial Cauchy slices. The reason is the effective physical Hamiltonian density $\mathcal{H}_{\Delta}(t,x)$ is not a conserved quantity. The only vacuum solution is the Minkowski solution. As a result, the vacuum solution for a massive object in the effective theory can only be achieved at $z =x-t= \pm \infty$, which reduces to the classical Schwarzschild solution. Lacking of non-trivial vacuum solution means the dust contribution will always appear except for some certain fixed isolated $t$'s. Thus the deparametrization is always well-defined in the system. 

\textbf{Remark 5:} The ODEs \eqref{ODE1} have a time-reflection symmetry with the following transformation of the fields:
\be\label{time_ref}
E^x(-z) \to - E^x(z), \; E^{\varphi}(-z) \to E^{\varphi}(z)\; K^x(-z) \to K^x(z)\; K^{\varphi}(-z) \to - K^{\varphi}(z) 
\ee
This symmetry can be used to define the white hole solution which can be glued asymptotically to the black hole solution as described in \cite{Han:2020uhb}.

\subsubsection{Approximate gluing to an effective static exterior solution}
Since our effective equations of motion (\ref{pde1})-(\ref{pde4}) hold for the whole spherical symmetric space time, a consistent exterior solution should solve (\ref{pde1})-(\ref{pde4}) as well at effective level. 
Beyond the classical theory we can still assume $x'(t) = 0$ and $\tau'(t) = 1$. The reason to chose this ansatz is that the surface of the star must follow a timelike geodesic of the exterior metric, where $x'(t) = 0$ is a timelike geodesic with metric (\ref{metric1}). 
The junction condition then becomes (assuming $\sigma_{ab} = 0$)
\begin{eqnarray}\label{new_junction}
    E^x(\tau,x)|_{x=x_s} = x_s^2 a(t)^2 \,, \qquad 2 E^{\varphi}(\tau,x)|_{x=x_s} = \frac{\partial_x E^x}{\sqrt{1-kx^2}} \Big|_{x=x_s},
\end{eqnarray}
which is similar to the classical one. 
This imposes the boundary condition at $x= x_s$ for the PDE system (\ref{pde1}-\ref{pde4}). At $t \ll 0$, the system approaches the classical regime, and assuming that it is approximately described by Einstein field equations in LTB spacetime we can set the initial value as the LTB vacuum solution. Under this approximation the boundary-initial value PDE system which can be solved numerically. 

\begin{figure}[H]
\centering
\includegraphics[width=9cm]{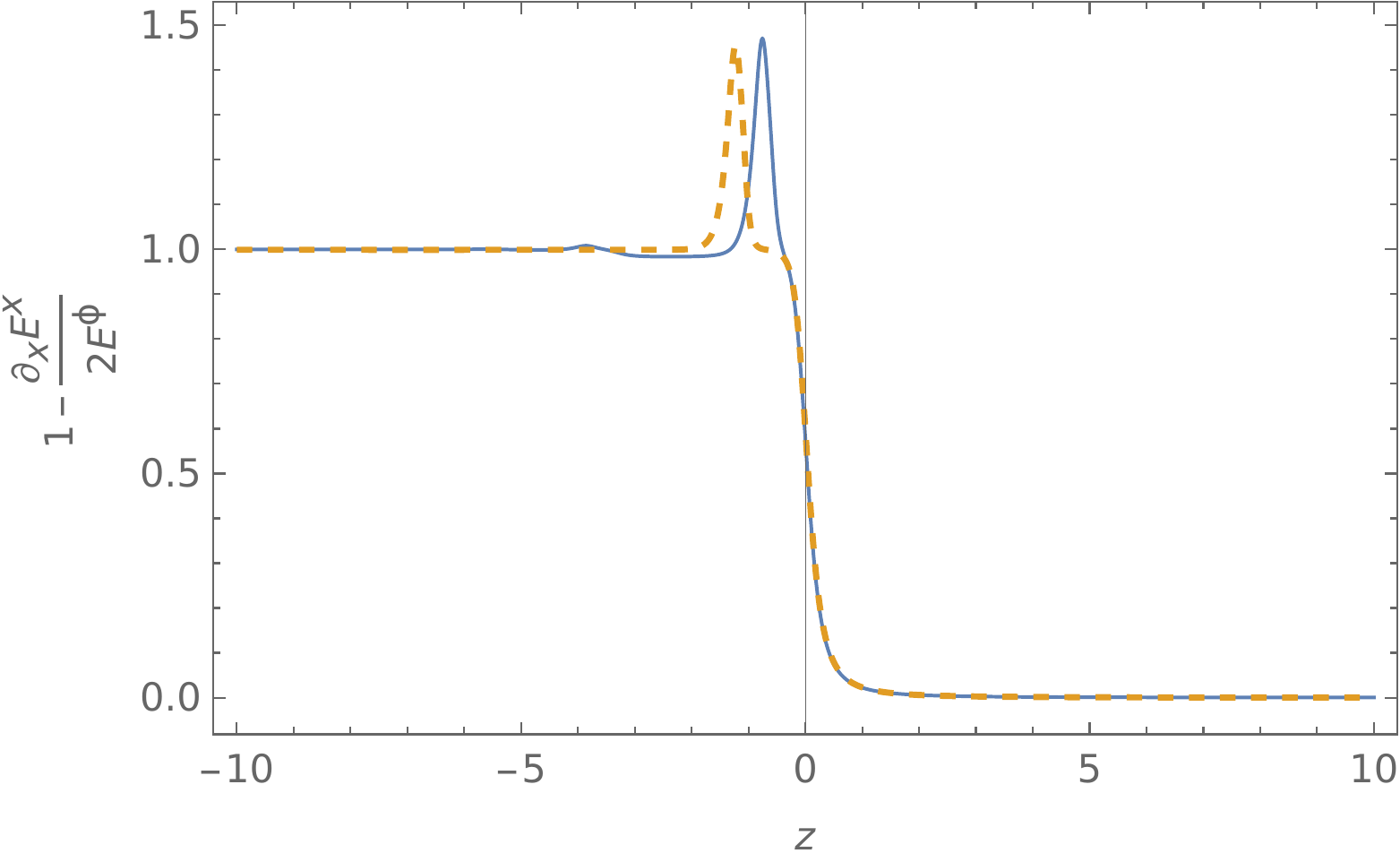} 
\caption{The evaluation of the LTB condition of the effective static exterior solution with mass $m=10$ (blue-solid line) and $m=1000$ (orange-dashed line) on the junction surface $x=x_s$ (Units are Planckian). The surface stress-energy tensor is given by $\sigma_{\mu \nu} dx^{\mu} dx^{\nu} = x_s a(t) (1-\frac{\partial_x E^x}{2 E^{\varphi}}) d\Omega^2 $.  The classical singularity lies at $z=0$. } 
\label{fig:ltb_sigma}
\end{figure}

In the case where the exterior solution has the  killing field $\partial_K$ given in \eqref{killing}, we have 
\begin{eqnarray}\label{new_junction_killing}
    E^x(z) = x_s^2 a^2(x_s-z) \,, \qquad 2 E^{\varphi}(z) = \frac{E^x{}'(z)}{\sqrt{1-kx_s^2}} = f(x_s) E^x{}'(z),
\end{eqnarray}
where the condition on the junction surface is transported to $z$ plane due to the killing field $\partial_K$, and $f(x_s):=\frac{1}{\sqrt{1-k x_s^2}}$. In this case, the metric \eqref{metrictz} becomes
\begin{equation}\label{metrictz1}
    ds^2 = -\lt( 1 - \frac{f(x_s)^2E^{x}{}'(z)^2}{4 E^x(z)}\rt) d \tau^2 + \frac{1}{\frac{4E^x}{f(x_s)^2E^{x}{}'(z)^2 } -1} d z^2 + E^x(z) d \Omega^2.
\end{equation}
Using the Friedmann equation \eqref{Friedmann from homogeneous reduction} we have
\begin{eqnarray}
   R'(z)^2 = \frac{(E^x{}'(z))^2}{4|E^x|}
    = \left(\frac{2 G m_s}{\sqrt{E^x}}- {k x_s^2 } \right) \left( 1 -\frac{\beta ^2 \Delta     }{E^x}\Big( \frac{2  G m_s}{ \sqrt{E^x}}-{k x_s^2} \Big) \right).
\end{eqnarray}
Thus the metric only depends on $R'(z)^2$.
In the region where $E^x(z)$ is monotonic, using $R := \pm \sqrt{E^x}$ which corresponds to the stage before or after the bounce, the above metric can be further rewritten as 
\begin{eqnarray}\label{metrictz1}
    ds^2& =& -\lt( 1 - f(x_s)^2R{}'(z)^2\rt) d \tau^2 + \frac{f(x_s)^2 }{1 - f(x_s)^2R{}'(z)^2} d R^2 + R^2 d \Omega^2 \\
    &=&- A(R) d \tau^2 + \frac{f(x_s)^2 }{A(R)} d R^2 + R^2 d \Omega^2,
\end{eqnarray}
with $A$ given by
\begin{equation}
    A(R) = 1 - f(x_s)^2 \left(\frac{2 G m_s}{|R|}- {k x_s^2} \right) \left( 1 - \frac{\beta^2 \Delta}{R^2} \Big( \frac{2  G m_s}{ |R|}- {k x_s^2} \Big) \right). 
\end{equation}
For $k=0$, we have $f(x_s)=1$ and
\begin{equation}
   A(R) = 1 - \frac{2 G m_s}{|R|}\left( 1 - \frac{\beta^2 \Delta}{R^2} \frac{2  G m_s}{ |R|} \right). 
\end{equation}
Note that now the metric \eqref{metrictz1} is well-defined for both before and after the bounce, as well as at the bounce. This metric has the same form as the one  obtained in \cite{Kelly:2020uwj,Lewandowski:2022zce}. However, here this metric is only defined locally before or after the bounce with a minimal value of $|R|_{\min} = \sqrt{E^x(z)}_{z=z_{bounce}}$, there is no extension to the regime  $R<R|_{\min}$.

\begin{figure}[h]
\includegraphics[width=0.47\linewidth]{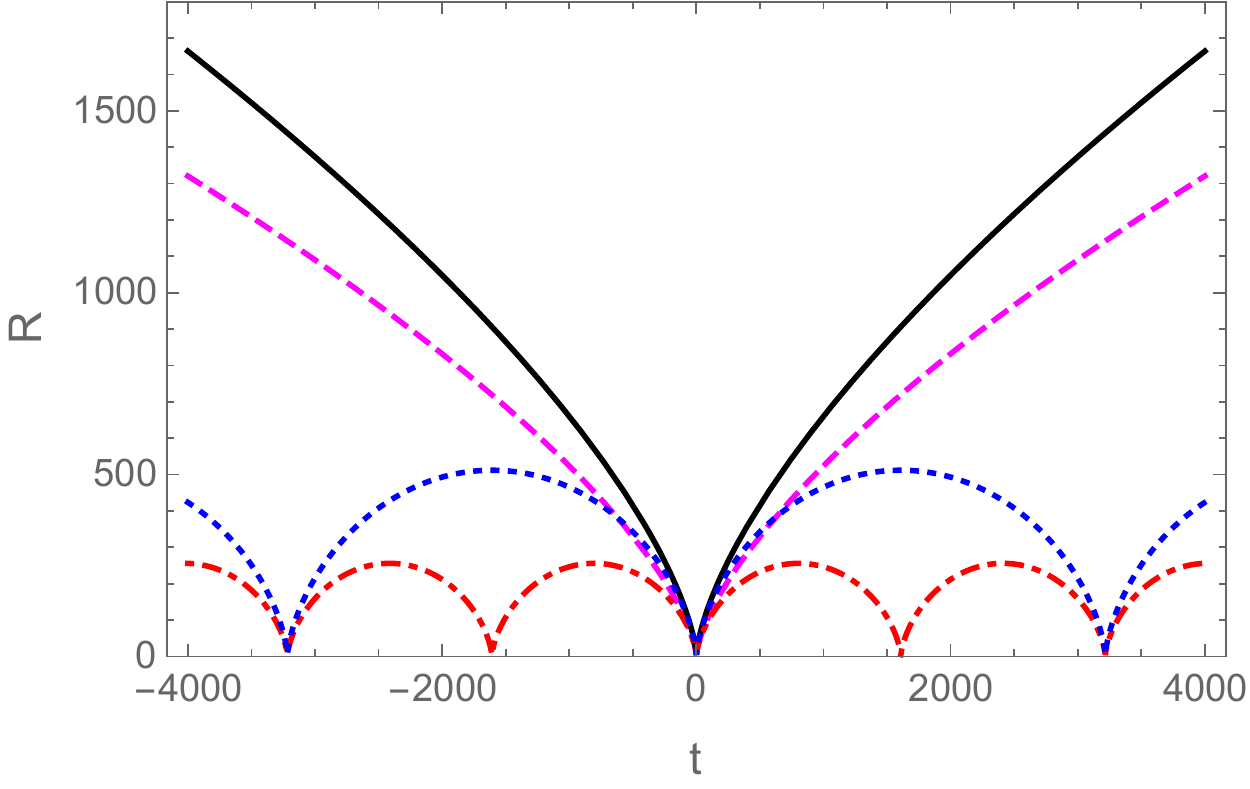}
\hspace{0.05\linewidth}
\includegraphics[width=0.45\linewidth]{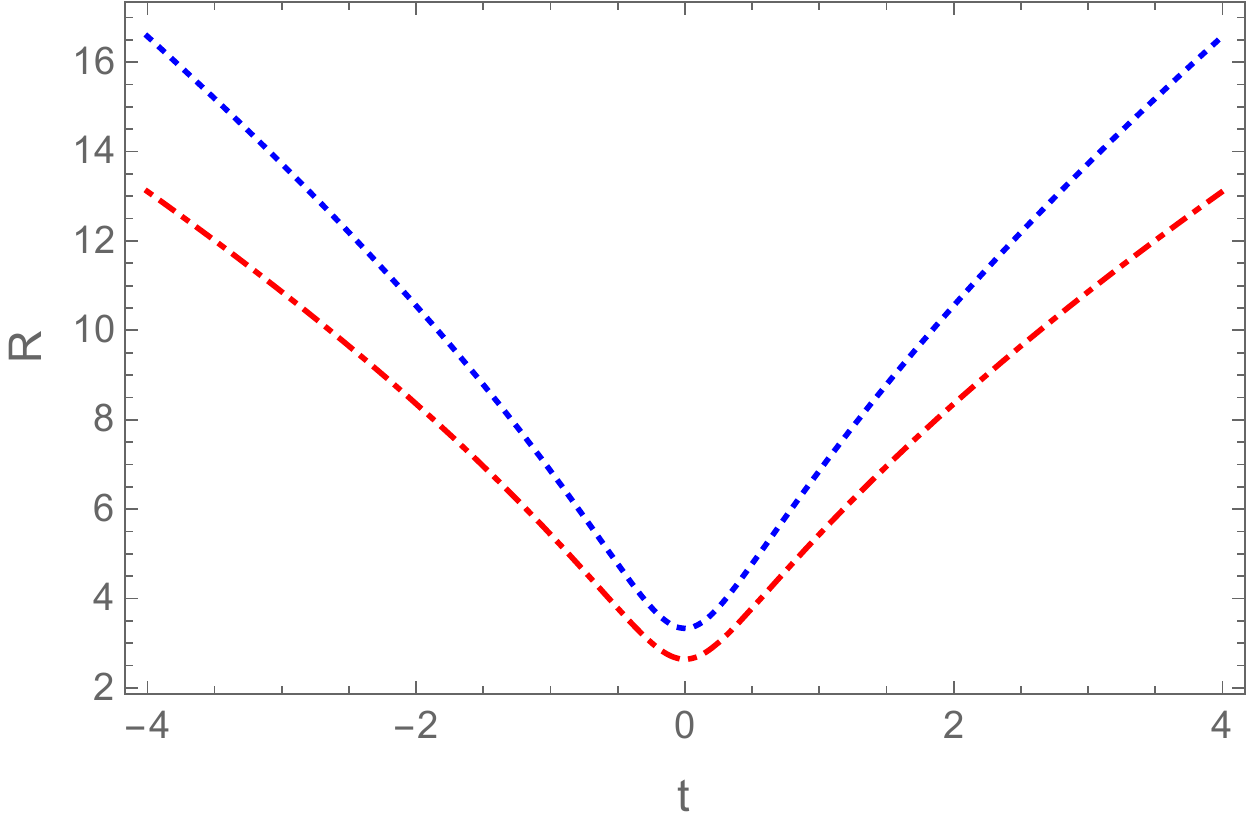}
\caption{In this plot, we compare the evolution of the physical radius for $k=0$  and $k=1$ with the mass given by $m_s=32$ and $m_s=64$. The left panel depicts these four cases, with  red dotdashed curve describing the case $(k=1,m_s=32)$, blue dotted curve for $(k=1,m_s=64)$, magenta dashed curve for $(k=1,m_s=32)$ and black solid curve for $(k=0,m_s=64)$.  On the right is the zoomed-in plot around the bounce at $t_b=0$ for the bound case only. 
}
\label{k1solution}
\end{figure}

It is important to note that (\ref{new_junction}) is exactly the first LTB condition on the junction surface. However, as one can see from \eqref{H_LTB}, the LTB condition is in general violated. More specifically, it is violated in the static solution given by \eqref{ODE1}. The violation is shown in Fig. \ref{fig:ltb_sigma}. As expected, the violation is strong in the quantum regime close to the classical singularity. Thus the junction condition with static exterior will lead to a non-trivial surface stress-energy tensor $\sigma_{\mu \nu} dx^{\mu}dx^{\nu} = \sqrt{E^x(\tau,x)}(1 -\frac{\partial_x E^x}{2 E^{\phi}(\tau,x)} )  d \Omega^2 \Big|_{x=x_s}$, which increases linearly with $\sqrt{E^x(\tau,x)}$ after the bounce., 
Such violation may relate to the fact that the static solution in effective dynamics contains a non-trivial dust mass distribution which is not compatible with the homogeneous interior.

\section{Numerical results of the effective homogeneous dust collapse}
\label{sec:numerics}
\renewcommand{\theequation}{4.\arabic{equation}}\setcounter{equation}{0}

In this section we present the numerical results of the dynamical evolution of the homogeneous dust collapse described by the effective dynamics for both marginally bound and the bound case.  We have already shown in Sec. \ref{sec:hamiltonian-and-conserved-quantity} that in the effective dynamics the classical singularity is resolved and replaced by a quantum bounce in the Planck regime. In our numerical simulations,  we carefully choose the initial conditions in the classical regime so that the bounce in the marginally bound case  takes place at time $t_b=0$. For the bound case we ensure that at least one of the bounces occurs at $t_b = 0$. In general, the \eqref{os metric} with solutions of the  effective equations of motion \eqref{reduced equations of motion} can be uniquely determined by three parameters, namely, the dust mass \eqref{dust mass}, the spatial curvature and the choice of the boundary surface. Therefore, in addition to quantum gravity effects, we will also investigate in some detail the impacts of these three parameters on the evolution of the homogeneous dust collapse, in particular, on the formation of the trapped surface during the contraction and the re-expansion  of the dust cloud. In the numerical results, we choose $\beta =0.2375$ based on the black hole  thermodynamics in LQG.

\begin{figure}[h!]
\centering
\includegraphics[width=0.46\linewidth]{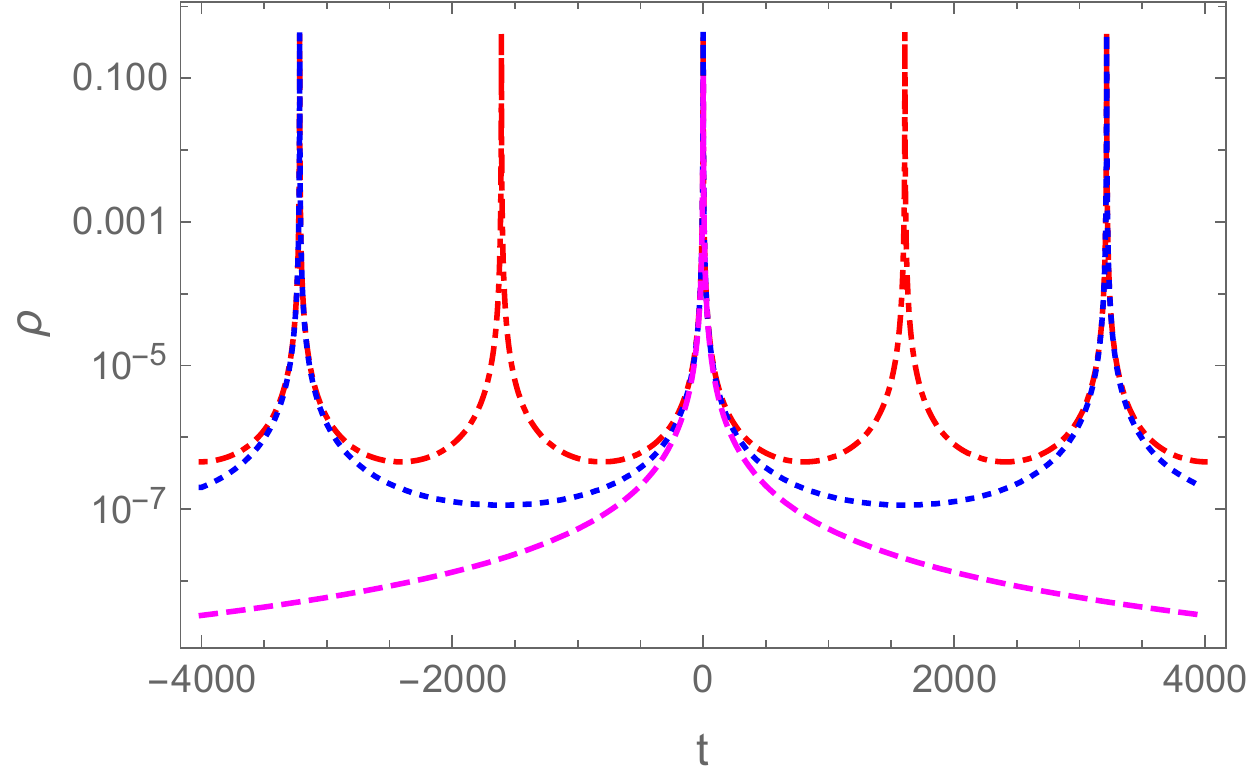}
\hspace{0.05\linewidth}
\includegraphics[width=0.46\linewidth]{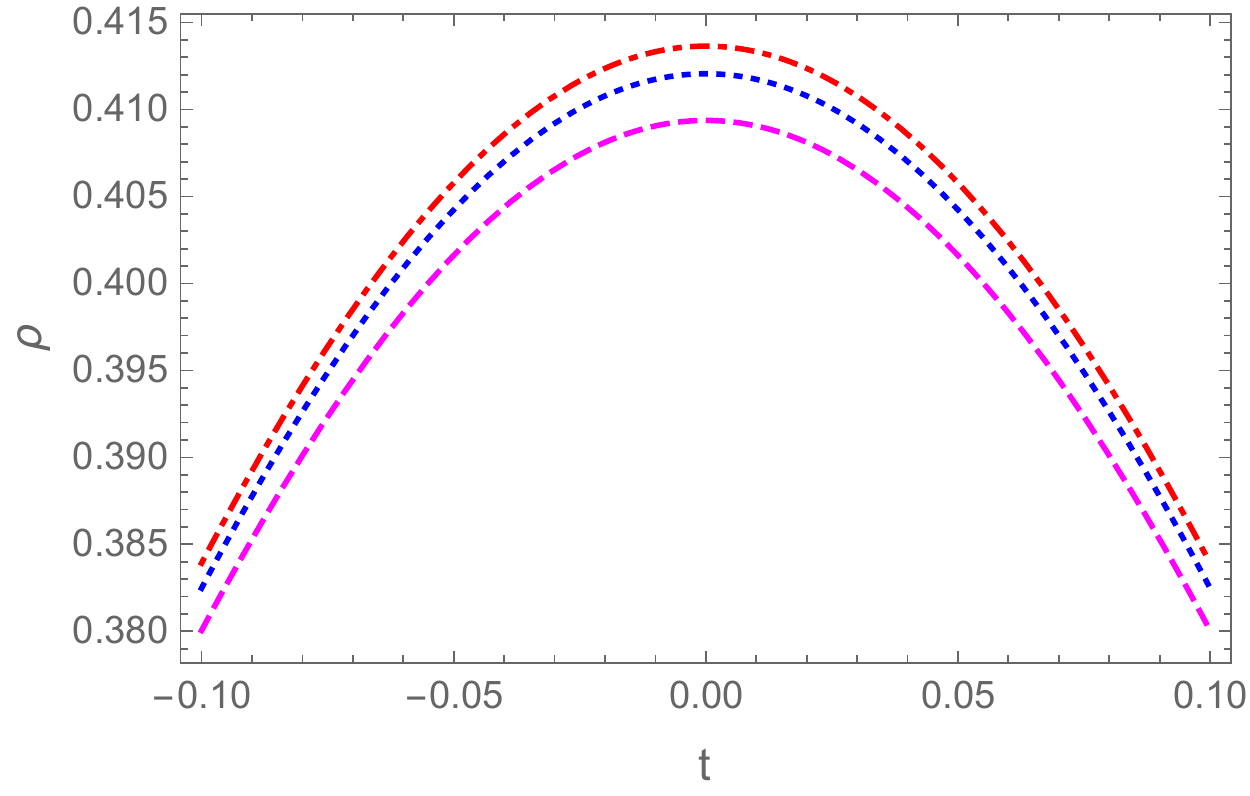}
\caption{In this plot we compare the dust energy density $\rho$ for $k=0$ (magenta dashed curve) and $k=1$ case, for $k=1$ we choose $m=32$ (red dot-dashed curve) and $m=64$ (blue dotted curve). In the right panel, we compare the energy density of these three cases near the bounce at $t=0$ which shows the difference of the maximum energy density among them.}
\label{k0k1rho}
\end{figure}

\begin{figure}[h!]
\centering
\includegraphics[width=0.46\linewidth]{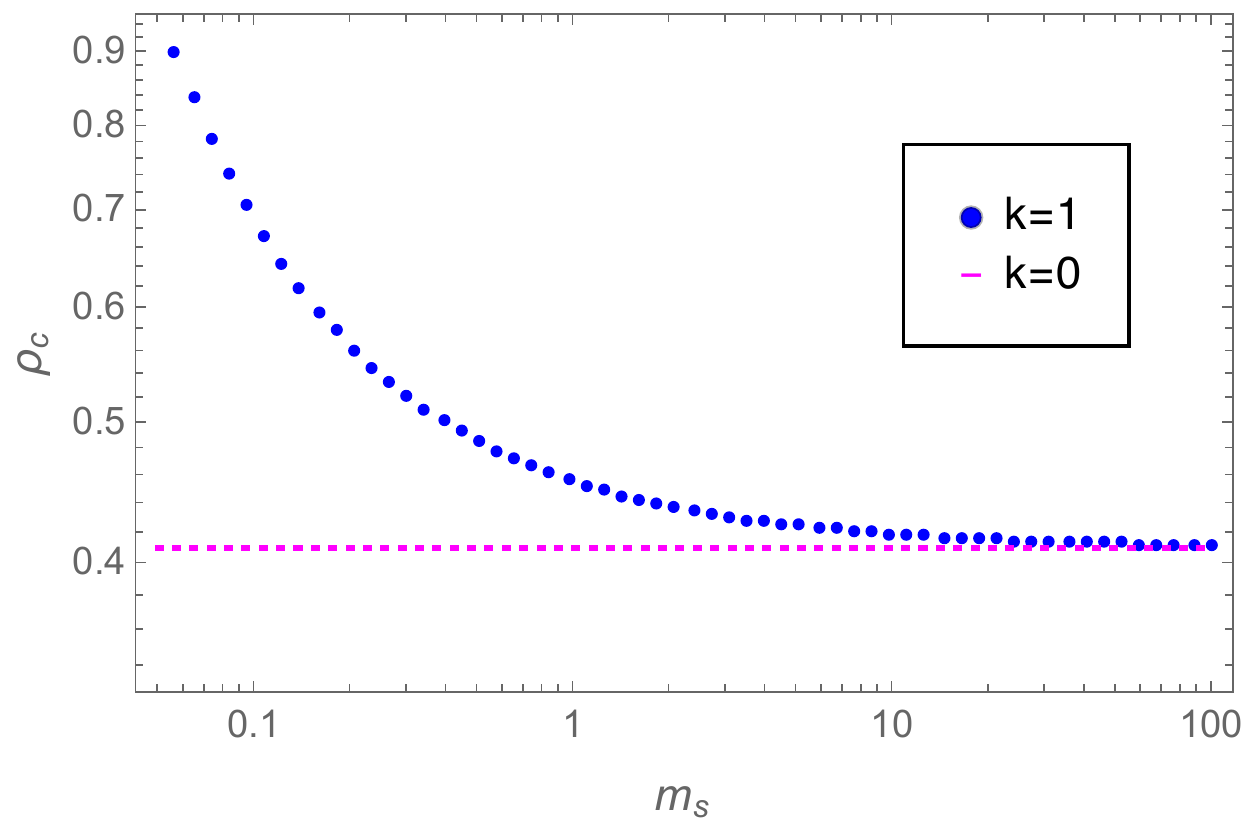}
\hspace{0.03\linewidth}
\includegraphics[width=0.46\linewidth]{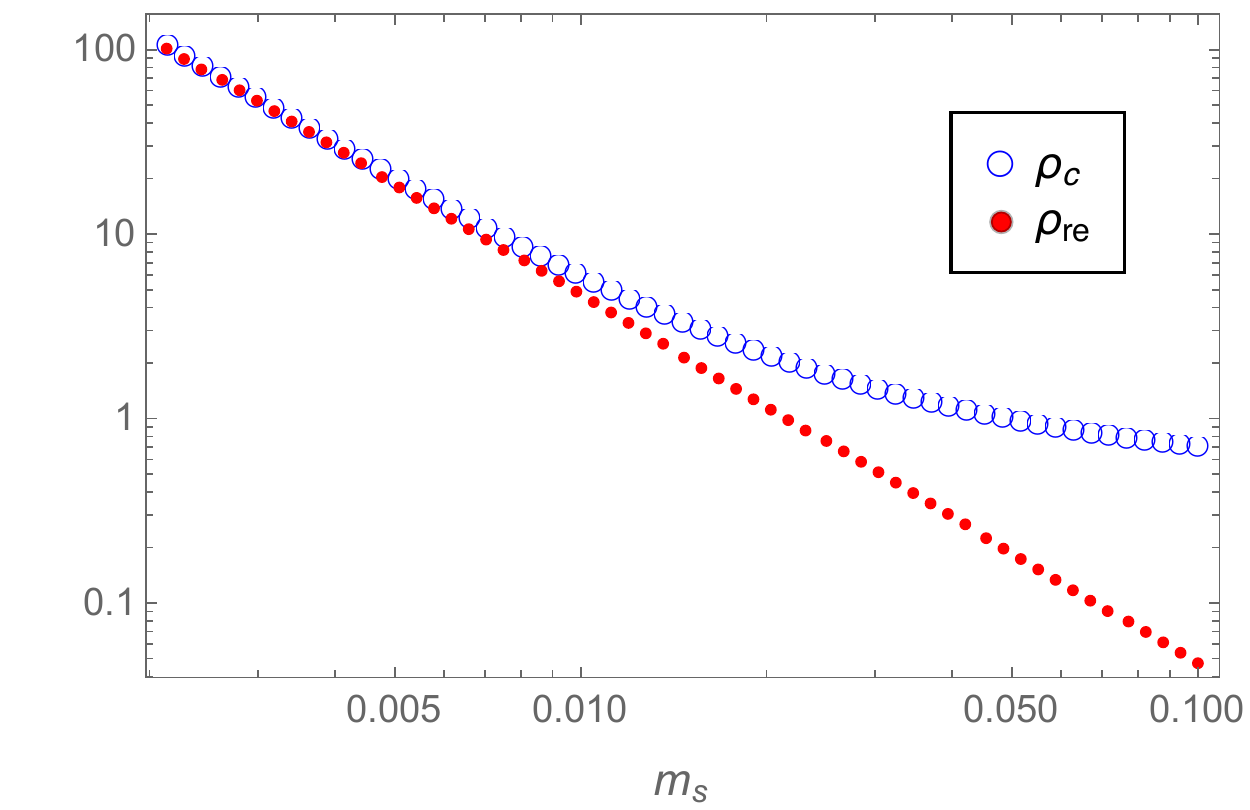}
\caption{In this plot we show explicitly how the mass of the dust cloud can affect the maximum energy density in two cases $k=0$ (magenta dashed line) and $k=1$ (blue dotted line) in the left panel. In the right panel, we show the change in the maximum energy density (blue circles) and the recollapse energy density (red disks) in the bound case for small dust mass. }
\label{k0k1rho1}
\end{figure}

In the following, we start with the qualitative evolution of the physical radius $R=x_s a$ of the dust cloud and investigate how it is affected by the dust mass and the spatial curvature. In Fig. \ref{k1solution}, we explicitly show the evolution of $R$ for the marginally bound and the bound case with the boundary surface chosen at $x_s=0.5$. For each case, two different dust masses are compared. The left panel depicts the behavior of $R$ over a long time for the bound case with mass $m_s=32$ (red dot-dashed curve) and $m_s=64$ (blue dotted curve) as well as the marginally bound case with the same mass, namely $m_s=32$ (magenta dashed  curve) and $m_s=64$ (black solid curve) while the right panels illustrate some details near the bounce at $t=0$. Since the physical radius evolves in a qualitatively same way for both marginally bound and the bound case, in the right panel we only show the details around the bounce in the bound case.   One can find from the figure the following properties of the homogeneous dust collapse in the effective dynamics.  Firstly, the classical singularity is generically resolved and replaced by a bounce for both marginally bound and the bound case. Secondly, for the marginally bound case, there exists only one single bounce which connects the collapsing phase with a re-expanding branch. The physical radius of the dust cloud reaches the minimal value at the bounce which increases with the dust mass. For the bound case, due to the spatial curvature, the dust cloud experiences identical cycles of contraction and expansion, mimicking the behavior of a pulsating star. The physical radius of the dust cloud increases at both the recollapse and the bounce point with the dust mass. Moreover, the period of the cyclic evolution of the dust cloud also increases with the dust mass. In particular, the period of cycles doubles when the dust mass doubles.

In Fig. \ref{k0k1rho}, we plot the change in the energy density for both cases with the parameters chosen the same as in Fig. \ref{k1solution}.  It turns out that the maximum energy density in the marginally bound case has an intrinsic value determined only by $\beta^2\Delta$ as given in (\ref{maximum energy density}). It would not be affected at all by the dust mass or the choice of the boundary surface. On the other hand, for the bound case, as expected in (\ref{med bound}), the maximum energy density also depends on the minimal physical radius of the dust cloud at the bounce which in turn is fixed by  $\beta$, $\Delta$ and the dust mass. In particular, as can be seen from \eqref{scale factor a}, when the dust mass $m_s\gg1$, $R_b = x_s a_b \sim m_s^{1/3}$. Therefore, the maximum energy density in the bound case decreases with the dust mass and tends to $\rho^{0}_\mathrm{max}$ in the marginally bound case. This feature is qualitatively  captured in the left panel of Fig. \ref{k0k1rho}. In particular,  the maximum energy densities of the bound case as seen from the red dot-dashed  and blue dotted curves are always larger than the maximum energy density of the marginally bound case represented by the magenta dashed curve. Meanwhile,  due to the effect of intrinsic curvature the blue dotted curve corresponding to $m_s=64$ has a lesser maximum energy density  than the red dot-dashed curve which corresponds to the dust mass $m_s=32$.

\begin{figure}[h!]
\centering
\includegraphics[width=7.7cm]{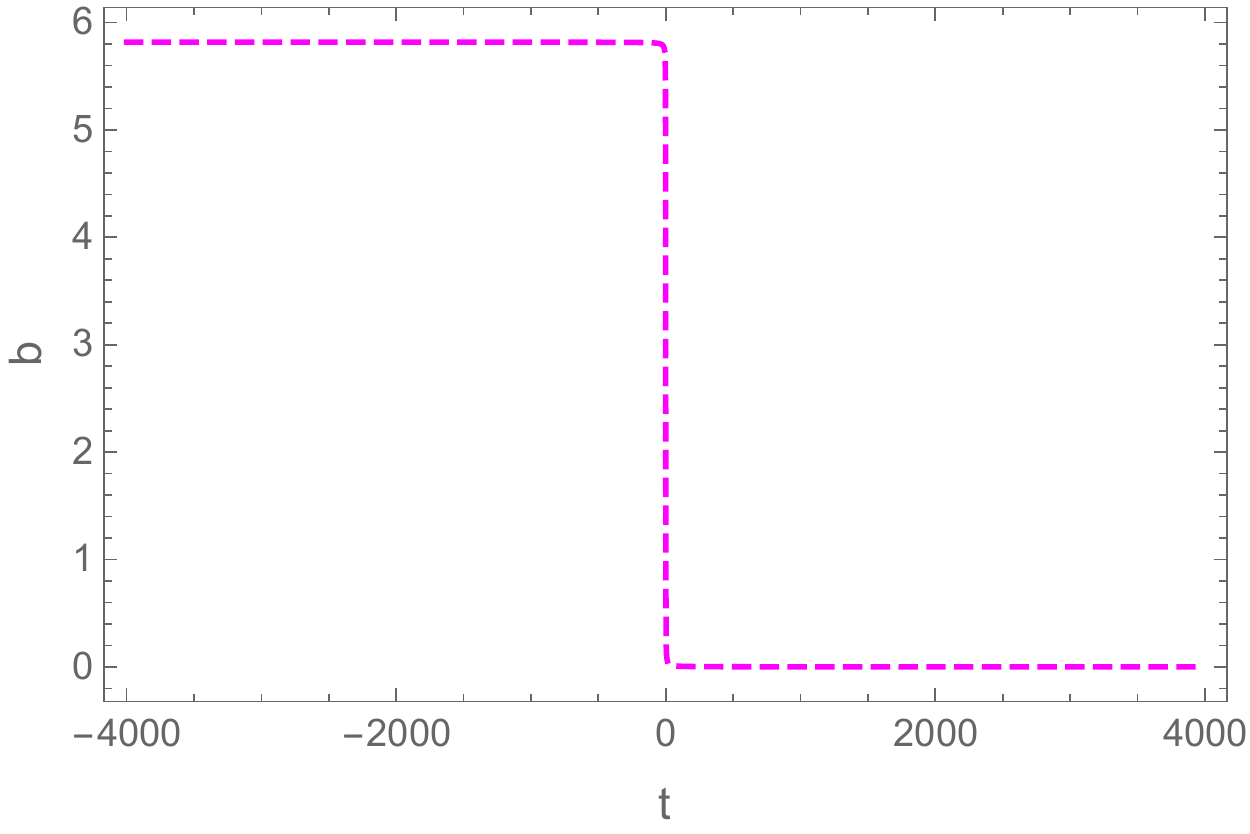}
\includegraphics[width=8cm]{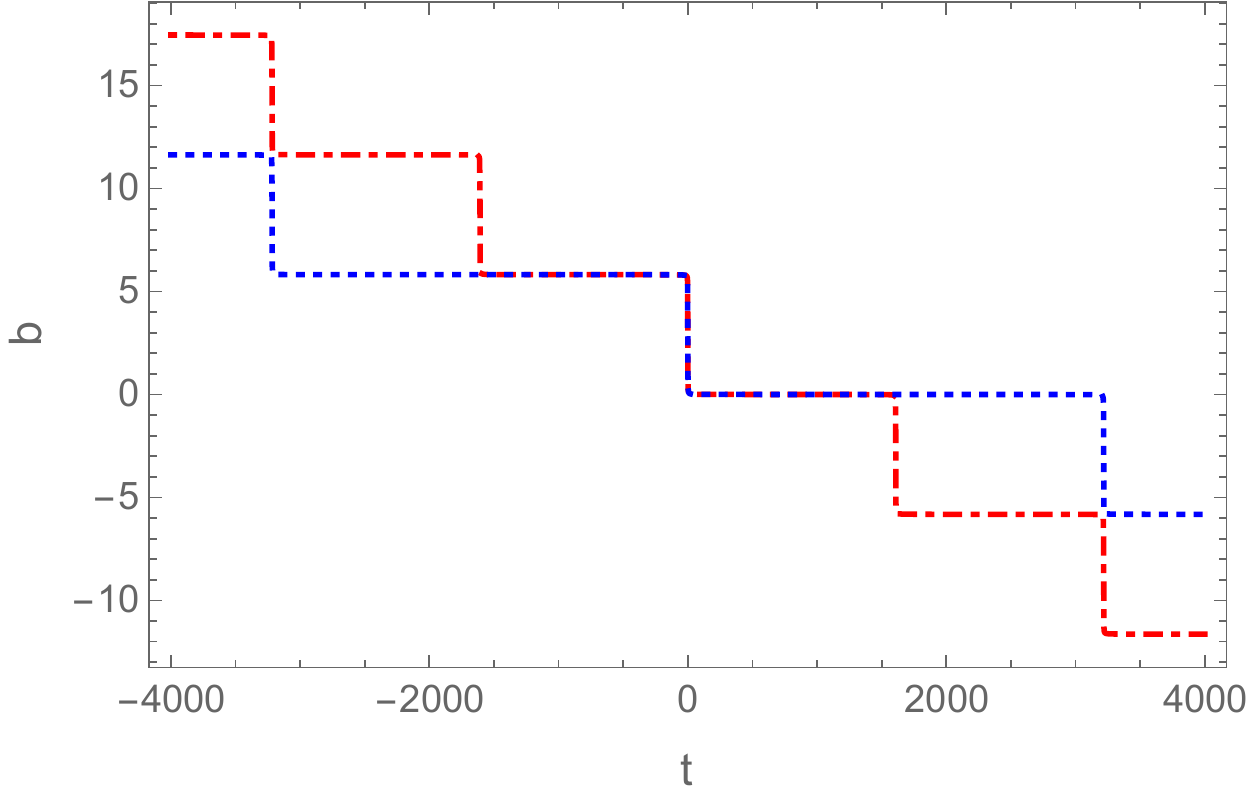}
\caption{In the figure, we show the evolution of the momentum $b$ in the marginally bound case in the left panel and also in the bound case in the right panel. The blue dotted curve corresponds to the case $(k=1,m_s=64)$ while the red dot-dashed curve to the case $(k=1,m_s=32)$.}
\label{k0solution}
\end{figure}

In Fig. \ref{k0k1rho1},   we show explicitly the dependence of the maximum energy density on the dust mass in two cases with a fixed boundary surface $x_s=0.5$. A different choice of the boundary surface would not change the value of $\rho^{0}_\mathrm{max}$ in the marginally bound case, neither it would change the way $\rho^{0}_\mathrm{max}$ depends on the dust mass in the bound case. As expected, the maximum energy density of the marginally bound case does not change with the dust mass as shown in the left panel of the figure while in the bound case the maximum energy density at the bounce point increases with a decreasing dust mass. When the dust mass increases, the shell recollapses at a larger value of the radius therefore making recollapse density negligible. In this case the density at the bounce in the marginally bound and bound cases approximate each other. For very small dust mass as depicted in the right panel of the figure, the difference between the maximum and the minimum energy densities tends to be small as compared with the maximum (minimum) energy density. This is why the blue circles appears to be overlapping with the red disks  at the small dust masses in the right panel.

\begin{figure}[h!]
\centering
\includegraphics[width=0.400\linewidth]{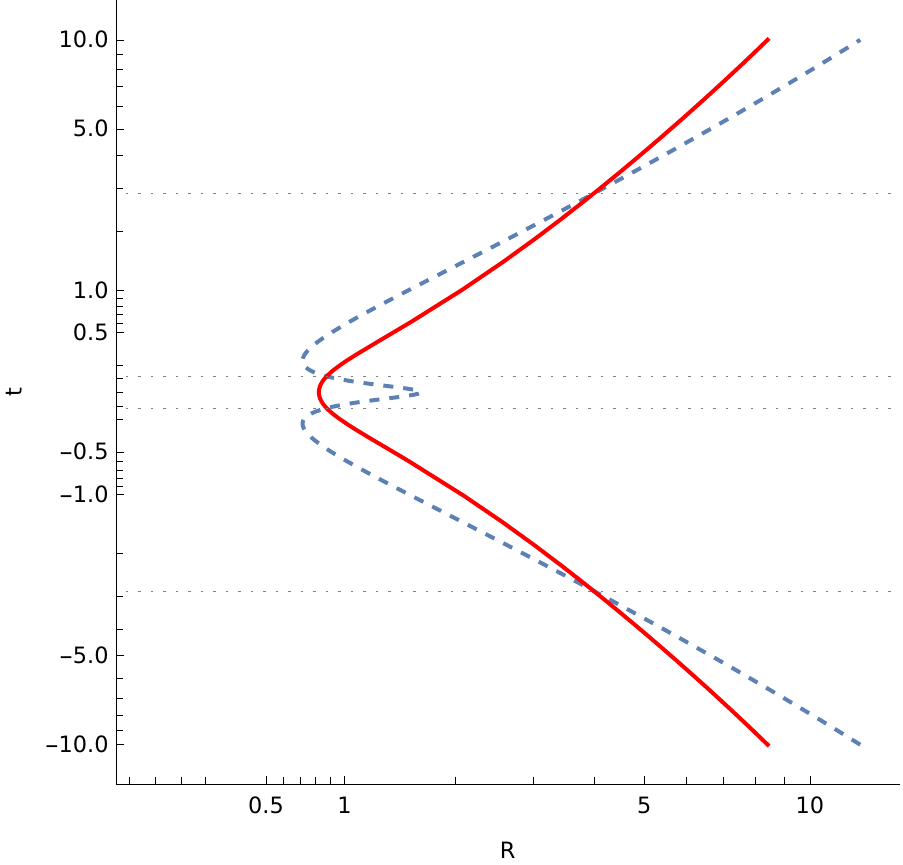}
\includegraphics[width=0.353\linewidth]{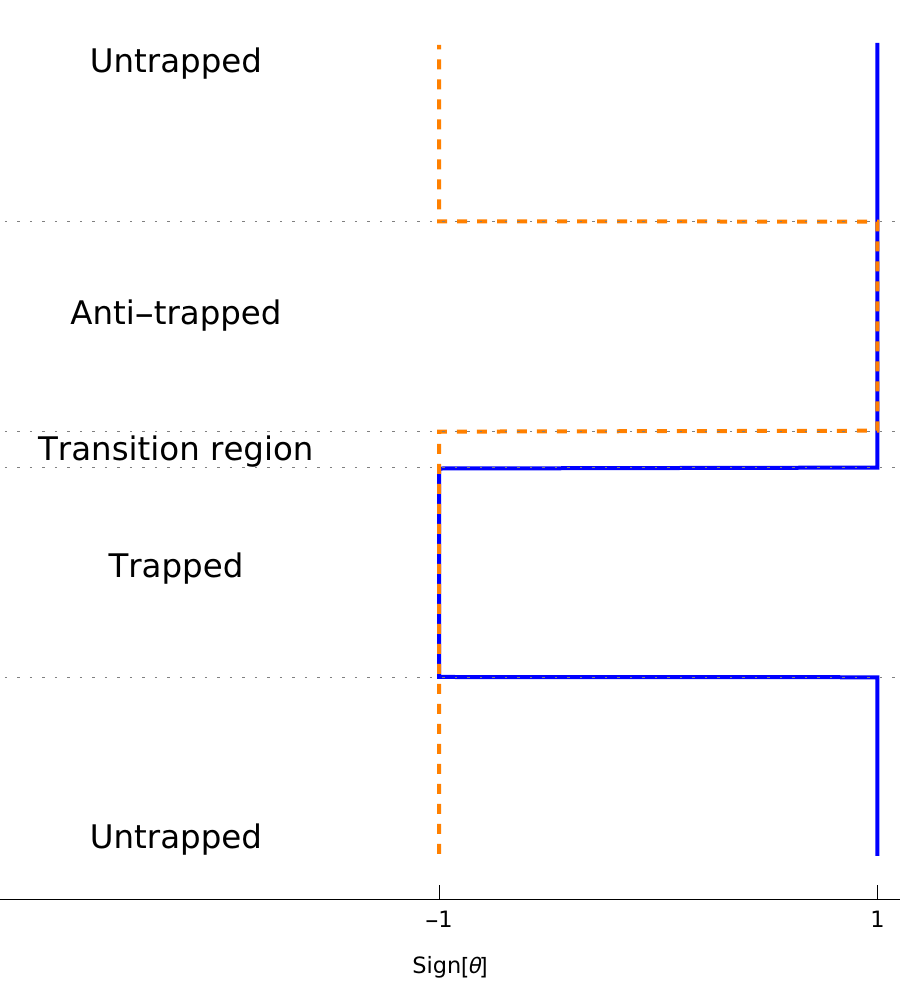}
\caption{The plot of the position of apparent horizon and the dust shell in bound case with $m_s =2$. The blue dashed line indicates the position of apparent horizon in the effective theory. %
The red lines indicates the evaluation of the dust shell under effective theory. The intersection between the apparent horizon and the dust shell separates the trapped and untrapped regions, which can be seen from the signature of null expansion on the right side (blue thick lines for $\theta_{+}$ and orange dashed lines for $\theta_{-}$).}
\label{horizon}
\end{figure}

In Fig. \ref{k0solution}, the qualitative behavior of the momentum $b$ is compared for the marginally bound and the bound case where we choose the dust mass $m_s=32, 64$ and the boundary surface $x_s=0.5$ as two examples. We find only the spatial curvature can affect the qualitative behavior of $b$. In particular, for the marginally bound case $k=0$, $b$ is confined within the range $b\in(0, \frac{\pi}{ \beta \sqrt{\Delta}})$. The classical limit can be recovered as $b\rightarrow 0$($b\rightarrow \frac{\pi}{ \beta \sqrt{\Delta}}$) in the collapsing (re-expanding) branch. A change in the dust mass would not affect the evolution of $b$ in this case. In contrast, for the bound case, the value of the momentum $b$ is not constrained in any finite range as the cyclic evolution of the dust cloud goes on. Since $b$ evolves monotonically when the matter content (such as the dust) satisfies the weak energy condition, it keeps decreasing during the forward evolution of the dust cloud in time. The classical limits are recovered near the recollapse points where $b$ takes the values $b=n\pi/\beta\sqrt\Delta$ with $n$ being any integers while the bounces take place at $b=l\pi/2\beta\sqrt\Delta$ with $l$ standing for any nonzero integers. In this way, the evolution of the momentum $b$ forms  a ladder structure in which there appears a plateau around the recollapse points $b=n\pi/\beta\sqrt\Delta$. We see the same  periods of the cycles in the bound case as in Fig. \ref{k1solution} and Fig. \ref{k0k1rho}. Since the dust mass can affect the periods of the cycles in the bound case, different masses can correspondingly change the duration of the plateau  as shown explicitly in the right panel of the figure.

\begin{figure}[h!]
\centering
\includegraphics[width=0.6\linewidth]{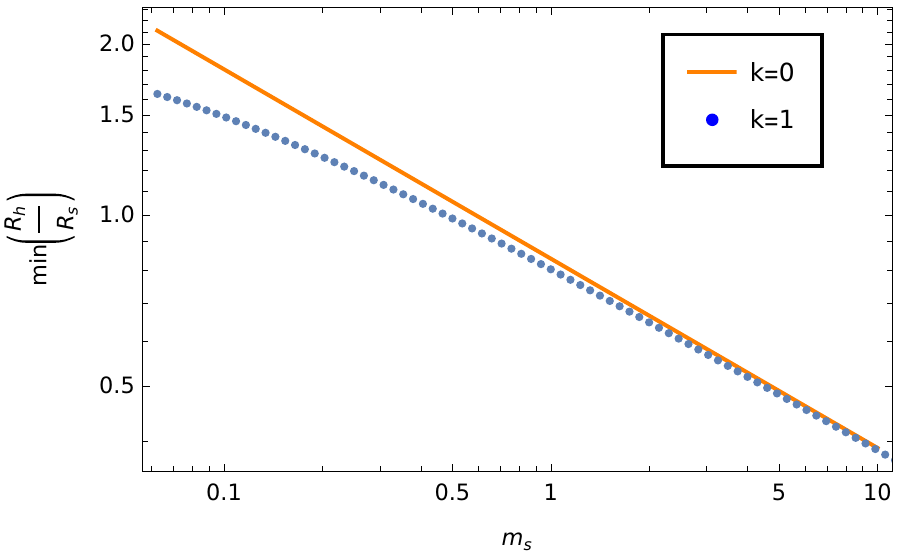}
\caption{In this figure, we numerically show the dependence of the minimum value of $R_h/R_s$  on the mass of the dust cloud in the cases of $k=0$ and $k=1$. When this minimal value is greater than unity, no horizon would form at the outermost dust shell during the non-singular evolution of the dust cloud. The threshold mass turns out to be  $M_{*}=0.5879$ for $k=0$ and $M_{*}=0.4756$ for $k=1$.}
\label{minimul_mass}
\end{figure}

Now let us discuss the formation of the trapped surfaces during the non-singular evolution of the homogeneous dust cloud in the effective dynamics. As discussed in Sec. \ref{sec:null-expansion-trapped-region-inner-horizon}, the outermost shell of the dust cloud becomes trapped as a result of its relative positioning against the apparent horizon which amounts to a marginally trapped surface. When the apparent horizon is located inside the outermost shell of the dust cloud, the latter becomes trapped and the dust cloud forms  a black/white hole. Therefore, it is of key importance to track the relative position of the outermost shell of the dust cloud with respect to the apparent horizon as shown explicitly in Fig. \ref{horizon}. In this representative example, we choose $x_s=0.5$, $m_s=2$ and $k=1$. Note we only pick one cycle with the bounce taking place at $t_b=0$.  Finally, we numerically compute the minimal value of the ratio between the apparent horizon and the location of the physical radius of outermost dust shell. The result is presented in Fig. \ref{minimul_mass}. As shown in the figure, this ratio only depends on the  mass of the dust cloud. For each case, there exists a threshold value of the dust mass. When the actual mass of the dust cloud falls below the threshold value,  no horizon would form during the non-singular evolution of the dust cloud. This actually sets up a lowest bound on the mass of the black hole which is formed by the collapse of the dust cloud. Besides, we also note that the spatial curvature does impact the specific value of the  threshold mass although its effect is limited.  

\section{Conclusions}
\label{sec:conclusion}
In this work, we applied the $\bar{\mu}$-scheme effective dynamics of the spherical symmetry reduced model to study the gravitational collapse for a homogeneous dust cloud. The model, having infinitely many physical DOFs, is developed based on the reduced phase space formulation of gravity coupled to Gaussian dust. The dust serves as both the reference field and the source of the gravitational collapse. Inside the dust cloud, the effective dynamics improves the classical Oppenheimer-Snyder (OS) model by resolving the singularity with a non-singular bounce, where the curvature is of Planckian  order. The effective dynamics from the model presented here for a homogeneous dust cloud reduces precisely to the effective dynamics of LQC with $K$-quantization based on using holonomies of the extrinsic curvature, indicating that the LQC effective dynamics for the spatially flat case lives as a subsector of the model presented here. Since the model presented in this work allows to consider the $k=0$ and $k=1$ case in a unified framework, we have also compared the properties of the two cases. Here we restate our  assumption  that the bound case only takes into account quantum geometric effects of extrinsic curvature via holonomies and the intrinsic curvature does not affect holonomies.  In  a former work by some of the authors  \cite{Giesel:2021dug} on the one hand  only the $k=0$ case was considered and on the other hand the LTB conditions were applied before quantization. It has been found that the spatial curvature can affect the qualitative dynamics of the evolution of the dust cloud. For $k=0$, the collapsing dust cloud bounces at a fixed  maximum energy density and then keeps expanding ever after. In this process, the momentum $b$ is confined within the range $b\in(0, \frac{\pi}{ \beta \sqrt{\Delta}})$ and monotonically increases. On the other hand, the evolution of the dust cloud in the bound case (with $k=1$) exhibits a richer dynamical properties. Firstly, due to the non-vanishing spatial curvature, the dust cloud experiences infinite cycles of contraction and expansion, mimicking the behavior of a pulsating star. Besides, both the maximum energy density at the bounce point and the minimum energy density at the recollapse point decrease with an increasing dust mass. In particular, for sufficient large values of the dust mass, the bounce energy density will merge into the same maximum energy density obtained in the marginally bound case and the gap between the maximum and the minimum energy densities also increase as the dust mass increases. 

Moreover, we find that the spatial curvature can also influence the threshold mass for the formation of the trapped surface at the outermost shell of the dust cloud during its gravitational collapse. To be specific, in the $k=0$ case, the analytical expression of the threshold mass is obtained and its exact value which is at the order of Planckian mass agrees with the one found in \cite{Giesel:2021dug} for the standard LQC quantisation where no covariant fluxes are involved.  In contrast, for the $k=1$ case, we can only obtain the numerical result of the threshold mass which turns out to be a slightly different value as compared with the one in the marginally bound case. In both cases, the bounce at the end of the gravitational collapse is symmetric in time reversal. Whenever there is a black-hole formed in the contracting phase of the dust cloud, it is always accompanied by an anti-trapped white-hole region in the expanding phase after the bounce which suggests that the white hole might be the final state of the black hole and  the dust is finally emitted by the white hole. The results obtained in this work also show that at least for the considered homegeneous collapse model and the $k=0$ case applying the LTB conditions in the classical theory and then quantizing yields the same threshold mass than if we first quantize, consider the corresponding effective model and implement the LTB conditions at the effective level.  Since the $k=1$ case was not considered in \cite{Giesel:2021dug} we cannot compare the results with the ones obtained here. The effective Hamiltonian used in this work connects to the $K$-quantization in LQC, because of the gauge fixing and the choice of basic variables discussed in Section \ref{review scheme}. The choice of basic variables affects the $\bar{\mu}$-scheme regularization in the effective Hamiltonian, and thus it affects the properties of the effective dynamics such as the conserved charges. The present choice results in that infinitely many charges of spatial diffeomorphisms are conserved. As the future investigation, it may be interesting to take into account the different choices, which might results in the dynamics with more conserved charges. 

Our results of the gravitational collapse and bounce is in favor of the black-hole-to-white-hole transition proposed and explored in e.g. \cite{Rovelli:2014cta11,DAmbrosio:2020mut,Bianchi:2018mml}. The effective dynamics here has the advantage of treating the black hole interior and exterior in a unified manner, and is aiming at a complete description of the non-singular black hole spacetime from the center to the infinity. The effective spacetime obtained here shares similarities with the proposal in e.g. \cite{Bianchi:2018mml}. However our effective description is still not complete, because the matching condition between inside and outside are not satisfied due to quantum effects. A detailed analysis of this issue is needed to assess whether this is because of the breakdown of the matching condition or the underlying scheme. 
In future work an investigation will be carried out to understand the region near the bounce, and possibly for this we will need a model involving inhomogeneous dust. To formulate such a model requires to extend the analysis on the LTB conditions at the effective level. For the homogeneous case considered here for the $k=0$ as well as the $k=1$ models the LTB conditions are preserved under the effective dynamics. As discussed in \cite{Bojowald:2008ja} and also briefly at the end of subsection \ref{sec:OSdustmodel} for an inhomogeneous model this is no longer given and thus one needs to carry over the LTB conditions consistently to such an effective model. An approach where the LTB conditions have been modified by functions depending on either one triad or extrinsic curvature variable depending on the chosen polymerization in the framework of effective techniques can be found in \cite{Bojowald:2008ja}. In future work we plan to investigate the stability LTB conditions further in order to see in addition to the work in \cite{Bojowald:2008ja} how the stability of the LTB conditions can be implemented in effective models \cite{GLRSWToAppear}. In addition presumably we might also take further effects from full LQG into account because of the strong quantum dynamical effects in the region around the bounce where the effective techniques considered here might not capture all properties of a given model. 

There are  proposals to glue the effective metric inside the dust cloud to the Vaidya solution outside the dust cloud, in order that the matching conditions are satisfied all the time. This proposal is used especially in the model where the effective dynamics inside and outside the dust cloud are treated separately (see e.g. \cite{Giesel:2021dug} and the references therein). The effective metric obtained here can also be glued to the Vaidya solution, as shown in Appendix \ref{Junction condition to Vaidya metric}. But this may become inconsistent in our approach, unless one is able to show that the Vaidya solution also satisfies the effective equations from ${\bf H}_{\Delta}$, because both the inside and the outside of dust cloud are governed by the same set of effective equations in the model presented here. It is still interesting to explore different effective solutions outside the dust cloud. The solution employed here assumes the timelike killing vector outside the horizon (so $E^x,E^{\varphi}$ are functions of $z=x-t$). Relaxing this killing symmetry might result in new dynamical solutions, which could satisfy the matching conditions even near the bounce. This will likely relate to the study of the inhomogeneous dust mentioned above, since the solution may correlate to nontrivial dynamics of the dust.

\begin{acknowledgements}

K. G., B.-F. Li and P.S are supported by DFG-NSF grants PHY-1912274 and 425333893. M.H acknowledges support by NSF grants PHY-1912278 and PHY-2207763. B.-F. Li also acknowledges support by the National Natural Science Foundation of China (NNSFC) with the grant No. 12005186. 
P.S. also acknowledges support by NSF grant PHY-2110207.  M.H. also acknowledges funding provided by the Alexander von Humboldt Foundation for his visit at the Friedrich-Alexander Universität Erlangen-Nürnberg. 

\end{acknowledgements}

\appendix
\section{Junction condition to the Vaidya metric}\label{Junction condition to Vaidya metric}

For the Vaidya metric
\begin{eqnarray}
\rmd s^2= - \left(1 - \frac{2G M(\tau,x)}{x} \right) \rmd \tau^2  -2 \rmd \tau \rmd x + x^2 \rmd \Omega^2,
\end{eqnarray}
eq. \eqref{junction_ltb} on junction surface $\Sigma$ determined by $x - x(\tau) =0$ becomes
\begin{eqnarray}
\gamma_{\mu \nu}^{+}  =-A^2 d\tau^2 + x(\tau)^2 d \Omega^2,\\
K_{\mu \nu}^{+} = \frac{1}{A} \left(-(x(\tau))^{-3}B d \tau^2 + C  d \Omega^2\right).
\end{eqnarray}
The matching condition \eqref{junction_condition} then implies
\begin{eqnarray}
A &=& \sqrt{\left(1 - \frac{2G M(\tau,x(\tau))}{x(\tau)} \right) + 2 x'(\tau)}, \\ 
B &=& -x(\tau) M(\tau,x(\tau)) \left(2 M^{(0,1)}(\tau,x(\tau))+3
   x'(\tau)+1\right) \\ 
   && +x(\tau)^2 \left(\left(3 x'(\tau)+1\right)
   M^{(0,1)}(\tau,x(\tau))+M^{(1,0)}(\tau,x(\tau))-x(\tau) x''(\tau)\right)+2
   M(\tau,x(\tau))^2, \notag\\
C &=& 2G M(\tau,x(\tau)) - x(\tau)(1+x'(\tau)).
\end{eqnarray}
Thus we obtain
\begin{eqnarray}
A \tau'(t) = 1, \qquad x(t) = x_s a(t),\qquad
B =0, \qquad C = A x_s a(t) \sqrt{1-k x_s^2},
\end{eqnarray}
which have the following solution
\begin{eqnarray}
\tau'(t) = \frac{x_s \sqrt{1-k x_s^2} a'(t )+1}{(1-k x_s^2) x_s^2 a'(t)^2-1}, \;\; M(\tau,x)|_{\Sigma} = \frac{1}{2} x_s^3 a(t) \left(1-k x_s^2\right)^{3/2} a'(t)^2,\\
x(t) = x_s a(t), \qquad \partial_x M(\tau,x)|_
{\Sigma} = \frac{1}{2} x_s^2 \left(1-k x_s^2\right)( a'(t)^2 + 2  a(t)  a''(t)).
\end{eqnarray}
The above solution gives the matching condition of the interior effective dynamics to exterior Vaidya spacetime.
In the $k=0$ case we obtain the junction condition given in \cite{Giesel:2021dug}.

\bibliographystyle{jhep}
\bibliography{refs}
\end{document}